\documentclass[aps,amsmath,amssymb,reprint,prd,superscriptaddress,nofootinbib]{revtex4-1}  
\usepackage{graphicx}  
\usepackage{dcolumn}   
\usepackage{bm}        
\usepackage{lineno}
\usepackage{hyperref}
\usepackage{amsfonts}
\usepackage{mathrsfs}
\usepackage{epsfig}
\usepackage{subfigure}

\def\aap{Astron.\ Astrophys.\ }

\def\apj{Astrophys.\ J.\ }
\def\apjl{Astrophys.\ J.\ Lett.\ }
\def\apjs{Astrophys.\ J.\ Supp.\ }

\def\mnras{Mon.\ Not.\ Roy.\ Astron.\ Soc.\ }

\def\prd{Phys.\ Rev.\ D\ }
\def\prl{Phys.\ Rev.\ Lett.\ }

\def\jcap{J.\ Cosmol.\ Astropart.\ Phys.\ }

\def\pasp{Publications\ of\ the\ Astronomical\ Society\ of\ the\ Pacific}

\def\nuflub8{\phi^\nu_B}
\def\nuflube7{\phi^\nu_{Be}}

\def\Rbr{\,R_{\mathrm{br}}}

\def\Rbr{\,R_{\mathrm{br}}}

\newcolumntype{p}{D{,}{\pm}{-1}}

\begin{document}

\title{Origin of the hardening in AMS-02 nuclei spectra at a few hundred GV}

\author{Jia-Shu Niu}
\email{jsniu@sxu.edu.cn}
\affiliation{Institute of Theoretical Physics, Shanxi University, Taiyuan 030006, China}
\affiliation{State Key Laboratory of Quantum Optics and Quantum Optics Devices, Shanxi University, Taiyuan 030006, China}
\affiliation{Collaborative Innovation Center of Extreme Optics, Shanxi University, Taiyuan, Shanxi 030006, China}

\begin{abstract}
  Many experiments have confirmed the spectral hardening in a few hundred GV of cosmic ray (CR) nuclei spectra, and 3 different origins have been proposed: the primary source acceleration, the propagation, and the superposition of different kinds of sources. In this work, the break power law has been employed to fit each of the AMS-02 nuclei spectra directly when the rigidity greater than 45 GV.  The fitting results of the break rigidity and the spectral index differences less and greater than the break rigidity show complicated relationships among different nuclei species, which could not been reproduced naturally by a simple primary source scenario or a propagation scenario. However, with a natural and simple assumption, the superposition of different kinds of sources could have the potential to explain the fitting results successfully. CR nuclei spectra from one single experiment in future (such as DAMPE) will provide us the opportunity to do cross checks and reveal the properties of the different kinds of sources.

\end{abstract}
\maketitle

\section{Introduction}

Cosmic-ray (CR) physics has entered a precision-driven era. More and more fine structures have been confirmed by a new generation of space-borne and ground-based experiments in recent years. For CR nuclei spectra, the most obvious fine structure is the spectral hardening at $\sim$ 300 GV, which was observed by ATIC-2 \cite{ATIC2006}, CREAM \cite{CREAM2010}, and PAMELA \cite{PAMELA2011}.

The space station experiment Alpha Magnetic Spectrometer (AMS-02), which was launched in May 2011, improve the measurement precision of the CR fluxes by an order of the systematics \cite{AMS2013}. Up to now, AMS-02 has released the spectra of different nuclei species, including the primary CR species: proton \cite{AMS02_proton}, helium (He), carbon (C), oxygen (O) \cite{AMS02_He_C_O}, neon (Ne), magnesium (Mg), and silicon (Si) \cite{AMS02_Ne_Mg_Si}; the secondary CR species: lithium (Li), beryllium (Be), and boron (B) \cite{AMS02_Li_Be_B}; the hybrid CR species: nitrogen (N) \cite{AMS02_N}. All these CR nuclei species show spectral hardening in the region of $100-1000$ GV, which confirms the previous observational results. Moreover, it shows that the secondary nuclei spectra harden even more than that of the primary ones at a few hundred GV, and the spectral index of N spectrum rapidly hardens at high rigidities and becomes identical to the spectral indices of primary He, C, and O CRs.

This spectral hardening phenomenon has been studied by many previous works. Generally speaking, the spectral hardening could come from: (i) the primary source acceleration (see, e.g., Refs. \cite{Ohira2011,Ptuskin2013,Korsmeier2016,Ohira2016,Boschini2017,Niu201801,Niu2019_dampe,Zhu2018,Niu2019,Yuan2019SCPMA,Niu2020,Boschini2020apj,Boschini2020apjs,Yuan2020jcap}); (ii) the propagation (see, e.g., Refs. \cite{Blasi2012,Tomassetti2012,Tomassetti2015apjl01,Tomassetti2015prd,Feng2016,Genolini2017,Jin2016CPC,Guo2018cpc,Guo2018prd,Liu2018,Niu2019,Boschini2020apj,Boschini2020apjs,Fornieri2020}); (iii) the superposition of different kinds of sources, such as different population of sources or of local and distant sources (see, e.g., Refs. \cite{Yuan2011,Vladimirov2012,Bernard2013,Thoudam2013,Tomassetti2015apjl02,Kachelriess2015,Jin2016CPC,Guo2016cpc,Kawanaka2018,Liu2018,Liu2019jcap,Qiao2019,Yang2019,Yue2019,Yuan2020}). Based on the galactic CR diffusion model, all these scenarios could provide good fits to specific data sets. No scenario has stood out yet.

  Although most of the previous works use a uniform injection spectrum for all the primary CR nuclei species or employ an independent injection spectrum for proton because of the proton-to-helium ratio anomaly \cite{Tomassetti2015apjl01}, some pioneering works (such as Refs. \cite{Yuan2019SCPMA,Boschini2020apj,Boschini2020apjs}) introduce different injection spectra for different primary CR nuclei species. Considering the different data sets and propagation models used in these works, it is natural that the injection spectral parameters of the same primary component are different. However, all these works show different primary CR injection spectral parameters for different primary CR species. This result should be paid more attention. Especially the break rigidities and the spectral index differences less and greater than the break of the species, these two kind of quantities could provide us some important information about the origin of the hardening directly.
  The observed CR spectra are physically produced by the injection spectra and the propagation process; even so, it is helpful to analyze all the observed/propagated AMS-02 nuclei spectra directly via a uniform method, which could provide us some robust conclusions about the primary source acceleration and the propagation of CRs.\footnote{The recent works which focus on the injection spectra (before propagation) based on specific propagation models can be found in Refs. \cite{Yuan2019SCPMA,Boschini2020apj,Boschini2020apjs} and related references therein.}

In the following, we first analyze the spectra in Section 2; discussions are shown in Section 3; conclusion and outlook are represented in Section 4.

\section{Analysis on the spectra}

Because we focus on the spectral hardening in a few hundred GV, the data points whose rigidity less than 45 GV are not used in this work, which are also affected by solar modulation and cannot be fitted by a simple break power law. When the rigidity is greater than 45 GV (up to a few thousand GV), all the nuclei spectra can be well fitted by a break power law or smooth break power law \cite{AMS02_proton,AMS02_He_C_O,AMS02_Ne_Mg_Si,AMS02_Li_Be_B,AMS02_N}. Considering the date precision of the AMS-02 data, it is unnecessary to employ a smooth factor to describe the spectra index transformation.\footnote{We also test the smooth break power law to fit the data, and it gives similar fitting results with slightly larger uncertainties on the parameters.}

Consequently, the following formula is used to describe each of the AMS-02 nuclei spectra (including the primary CR species: proton, He, C, O, Ne, Mg, and Si; the secondary CR species: Li, Be, and B; the hybrid CR species: N) when the rigidity is greater than 45 GV:
\begin{equation}
  F^{\mathrm{i}}(R) =  N^{\mathrm{i}} \times \left\{ \begin{array}{ll}
                                                       \left( \dfrac{R}{\Rbr^{\mathrm{i}} } \right)^{\nu^{\mathrm{i}}_{1}}  & R \leq \Rbr^{\mathrm{i}}\\
                                                       \left( \dfrac{R}{\Rbr^{\mathrm{i}} } \right)^{\nu^{\mathrm{i}}_{2}} & R > \Rbr^{\mathrm{i}} 
\end{array}
\right.,
\label{eq:bpl}
\end{equation}
where $F$ is the flux of CR, $N$ is the normalization constant, and $\nu_{1}$ and $\nu_{2}$ are the spectral indexes less and greater than the break rigidity $\Rbr$, and $\mathrm{i}$ denotes the species of nuclei.  The errors used in our fitting are the quadratic sum of statistical and systematic errors.

The Markov Chain Monte Carlo (MCMC) algorithm is employed to determine the posterior probability distribution of the spectral parameters belonging to different CR nuclei species.\footnote{The {\sc python} module {\tt emcee} \cite{emcee} is employed to perform the MCMC sampling. Some such examples can be referred to Ref. \cite{Niu201801,Niu201802,Niu2019} and references therein.}
The best-fit values and the allowed intervals from 5th percentile to 95th percentile of the parameters $\nu_1$, $\nu_2$, $\Rbr$, and $\Delta \nu \equiv \nu_2 - \nu_1$ are listed in Table \ref{tab:spectra_params}, together with the reduced $\chi^2$ of each fitting.\footnote{The information of the parameter $N$ is not listed in the table, which is not important in the subsequent analysis. The information of $\Delta \nu \equiv \nu_2 - \nu_1$ is derived from that of $\nu_1$ and $\nu_2$.} The best-fit results and the corresponding residuals of the primary, the secondary, and the hybrid CR species are given in Figure \ref{fig:pri_spectra}, \ref{fig:sec_spectra}, and \ref{fig:hyb_spectra} of the Appendix, respectively.

\begin{table*}
  \caption{The fitting results of the spectral parameters for the different nuclei species. Best-fit values and allowed 5th to 95th percentile intervals (in the square brackets) are listed for each of the parameters. }
\begin{tabular}{c|cccc|c}
\hline\hline
Species  &$\nu_1$  &$\nu_2$  &$\Rbr\ \mathrm{(GV)}$  &$\Delta \nu$  &$\chi^2 / \mathrm{d.o.f}$    \\
\hline
  proton          &-2.815\ [-2.823, -2.806]  &-2.71\ [-2.76, -2.62]  &379\ [300, 544]  &0.10\ [0.06,0.19]  & 1.21/27 = 0.045 \\
  Helium          &-2.725\ [-2.733, -2.715]  &-2.62\ [-2.65, -2.56]  &331\ [281, 448]  &0.10\ [0.07,0.16]  & 2.65/28 = 0.095\\
  Carbon          &-2.74\ [-2.76, -2.72]  &-2.64\ [-2.68, -2.59]  &202\ [148, 299]  &0.10\ [0.05,0.15]  & 5.26/28 = 0.188 \\
  Oxygen          &-2.696\ [-2.712, -2.680]  &-2.49\ [-2.63, -2.27]  &664\ [488, 964]  &0.21\ [0.07,0.43]  & 1.91/28 = 0.068 \\
  Neon            &-2.74\ [-2.76, -2.72]  &-2.33\ [-2.61, -1.98]  &670\ [405, 995]  &0.41\ [0.13,0.76]  & 6.01/27 = 0.222 \\
  Magnesium       &-2.74\ [-2.76, -2.72]  &-2.61\ [-2.79, -2.31]  &410\ [287, 978]  &0.13\ [-0.06,0.42]  & 4.68/27 = 0.173 \\
  Silicon         &-2.71\ [-2.73, -2.69]  &-2.79\ [-3.24, -2.51]  &922\ [491, 988]  &-0.08\ [-0.53,0.21]  & 7.21/27 = 0.267 \\  
\hline
  Lithium         &-3.18\ [-3.20, -3.10]  &-2.98\ [-3.01, -2.72]  &123\ [112, 351]  &0.20\ [0.14,0.41]  & 22.51/27 = 0.834 \\
  Beryllium       &-3.13\ [-3.16, -3.08]  &-2.95\ [-3.06, -2.77]  &199\ [173, 438]  &0.17\ [0.04,0.34]  & 18.29/27 = 0.677 \\
  Boron           &-3.10\ [-3.13, -3.07]  &-2.84\ [-2.96, -2.66]  &275\ [194, 422]  &0.26\ [0.14,0.44]  & 11.42/27 = 0.430 \\
\hline
  Nitrogen        &-2.93\ [-2.95, -2.87]  &-2.66\ [-2.70, -2.34]  &208\ [188, 454]  &0.27\ [0.21,0.56]  & 10.96/27 = 0.406 \\
\hline  
\end{tabular}
\label{tab:spectra_params}
\end{table*}

Generally speaking, the $\chi^2$s of primary CR species are smaller than the other 2 types of species, which are caused by the dispersion of the data points (especially in high rigidity region) in the latter cases. 

\begin{figure*}
  \centering
  \includegraphics[width=0.8\textwidth]{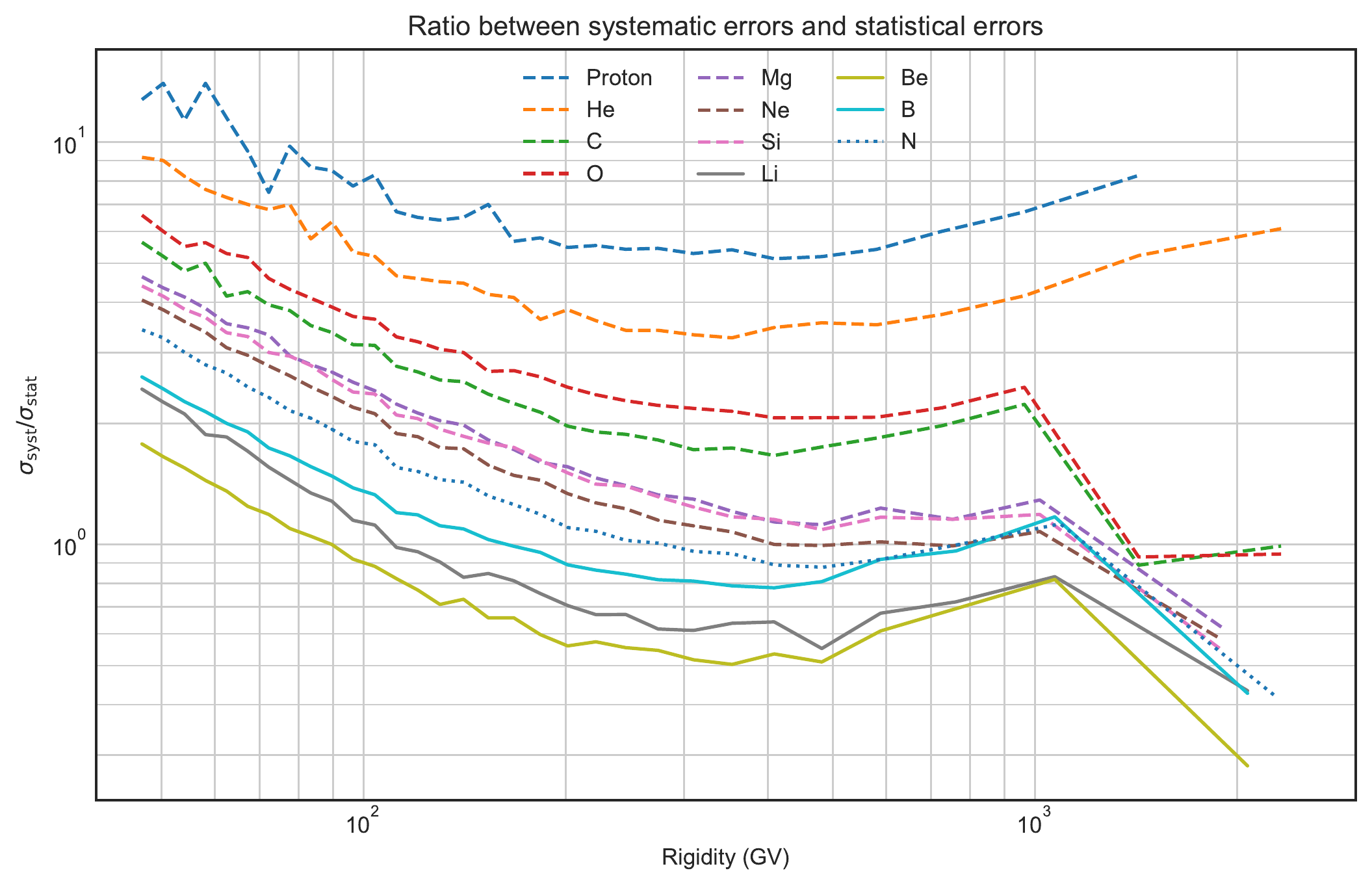}
\caption{Ratio between systematic errors and statistical errors $\sigma_{\mathrm{syst}}/\sigma_{\mathrm{stat}}$ with the variation of rigidity for different species. The primary CR species are represented in dashed lines, the secondary CR species are represented in solid lines, and the hybrid CR species is represented in dotted line.}
  \label{fig:data_errors}
\end{figure*}

 Here, one should note that the $\chi^2/\mathrm{d.o.f}$ of the best-fit result for the primary species (especially for proton, helium and oxygen) are much smaller than 1.0, which indicates an improper treatment of the data errors in the fitting process. In AMS-02 data \cite{AMS02_proton,AMS02_He_C_O,AMS02_Ne_Mg_Si,AMS02_Li_Be_B,AMS02_N}, we find that the systematic errors are always the dominating parts (see in Figure \ref{fig:data_errors}), which will lead to smaller $\chi^2$s if we ignore the energy correlations for them. Figure \ref{fig:data_errors} shows the ratio between the systematic errors ($\sigma_{\mathrm{syst}}$) and statistical errors ($\sigma_{\mathrm{stat}}$)  with the variation of rigidity. It is clear that the species who have the largest 3 $\sigma_{\mathrm{syst}}/\sigma_{\mathrm{stat}}$ values (proton, helium and oxygen) correspond to the smallest 3 $\chi^2 / \mathrm{d.o.f.}$ values in Table \ref{tab:spectra_params}. In such case, we need the correlation matrix of systematic errors of AMS-02 data if we want reasonable $\chi^2 / \mathrm{d.o.f.}$s for the fitting results. 
Unfortunately, the AMS-02 collaboration does not provide these correlation matrix of systematic errors. Consequently, the values of $\chi^2/\mathrm{d.o.f}$ in Table \ref{tab:spectra_params} does not have the absolute meanings of goodness-of-fit. Further data analysis needs more information about the systematic errors, and some detailed discussions about this topic can be referred to in Refs. \cite{Derome2019,Weinrich202001,Heisig2020}.

\section{Discussions}

\begin{figure*}
  \centering
  \subfigure[]{
  \centering
  \includegraphics[width=0.47\textwidth]{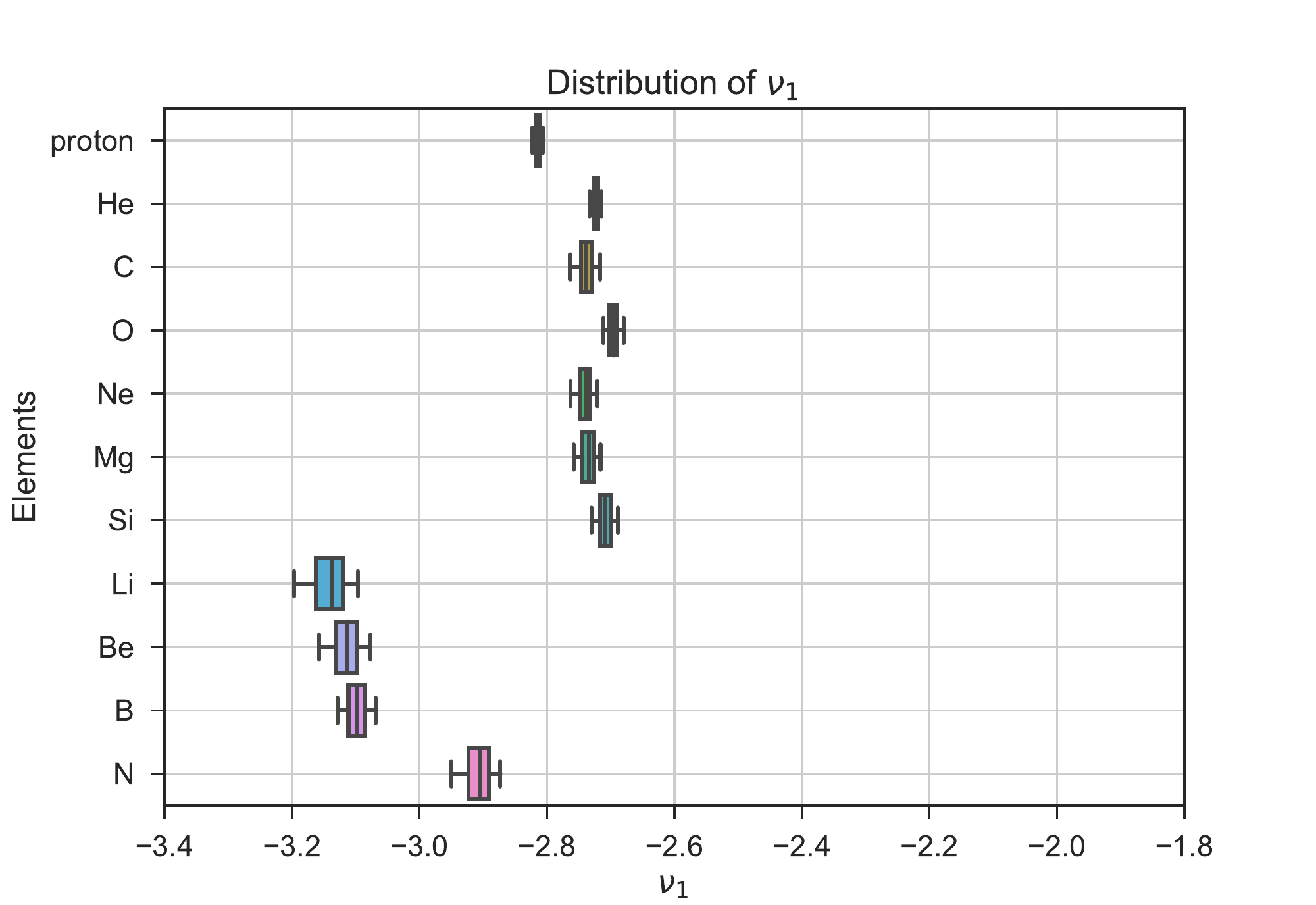}
  }
  \subfigure[]{
  \centering
  \includegraphics[width=0.47\textwidth]{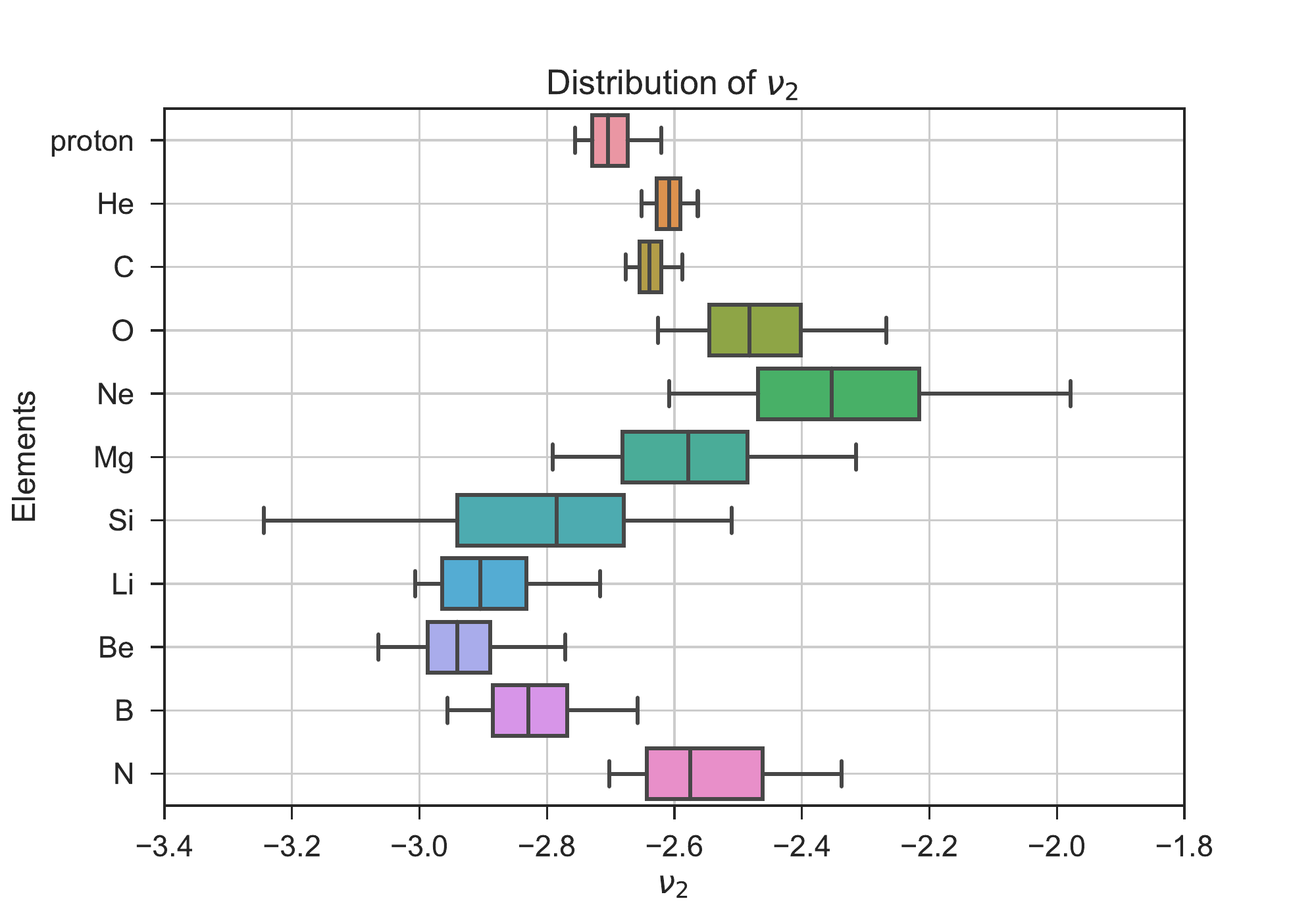}
}
\subfigure[]{
\centering
\includegraphics[width=0.47\textwidth]{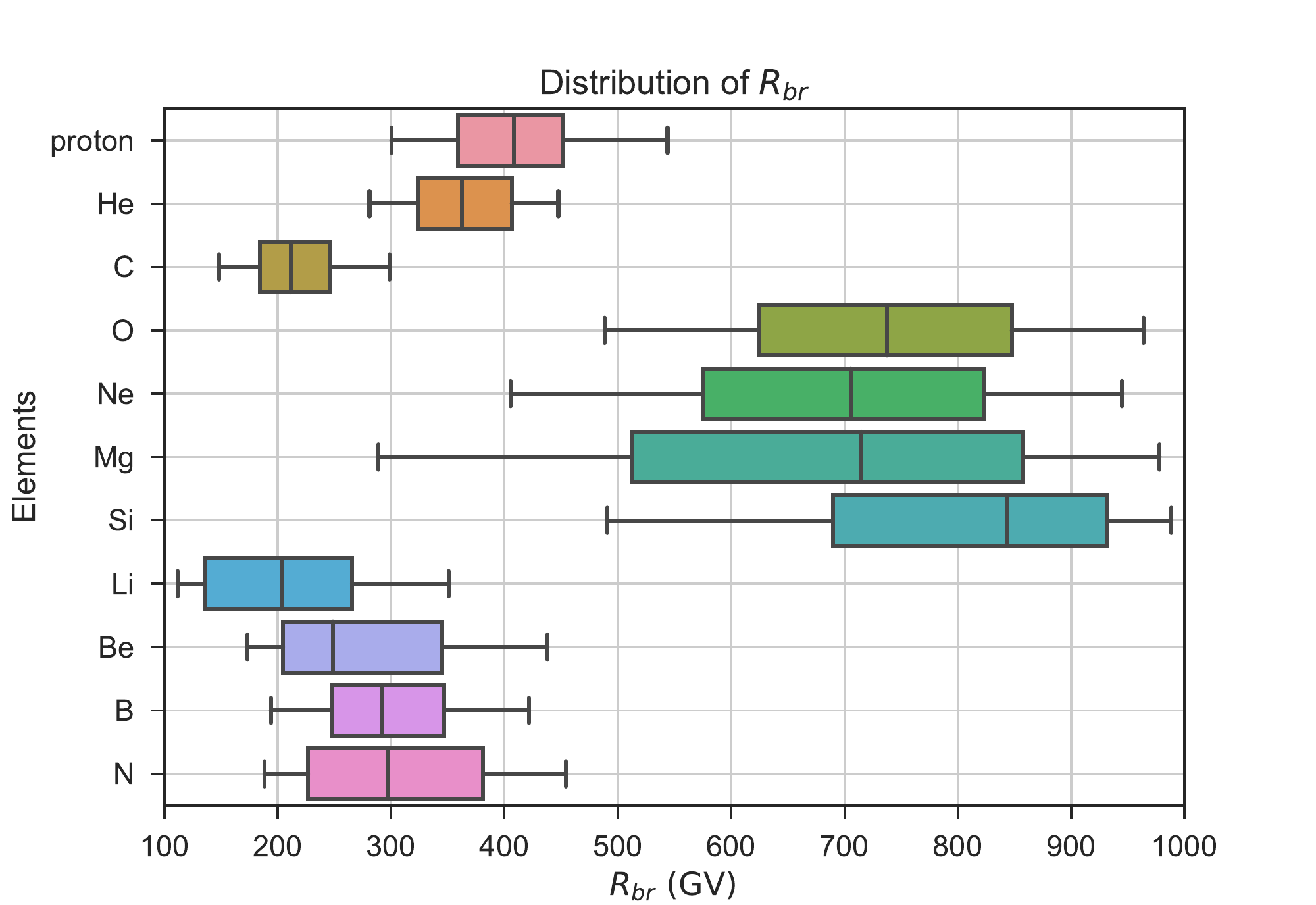}
}
\subfigure[]{
  \centering
  \includegraphics[width=0.47\textwidth]{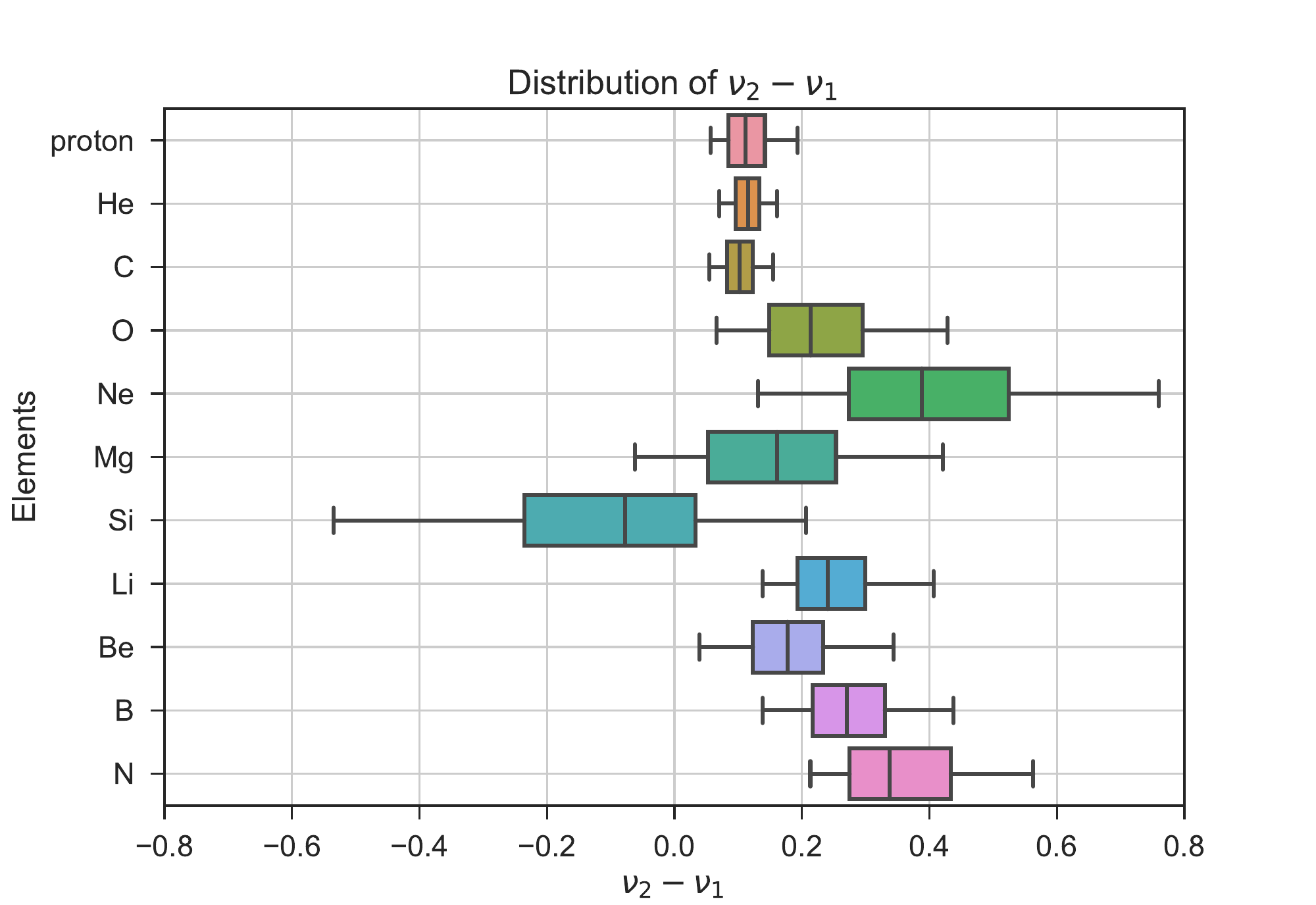}
}

\caption{Boxplots for $\nu_1$, $\nu_2$, $\Rbr$, and $\nu_2-\nu_1$. The band inside the box shows the median value of the dataset, the box shows the quartiles, and the whiskers extend to show the rest of the distribution which are edged by the 5th percentile and the 95th percentile.}
  \label{fig:spectra_params}
\end{figure*}

In order to get a clear representation of the fitting results, we use boxplot\footnote{A box plot or boxplot is a method for graphically depicting groups of numerical data through their quartiles. In our configurations, the band inside the box shows the median value of the dataset, the box shows the quartiles, and the whiskers extend to show the rest of the distribution which are edged by the 5th percentile and the 95th percentile.} to show all the distributions of $\nu_1$, $\nu_2$, $\Rbr$ and $\Delta \nu$ in Figure \ref{fig:spectra_params}.

In the subfigure (a) of Figure \ref{fig:spectra_params}, it is obvious that the values of $\nu_1$ can be divided into 3 groups, which correspond to the primary, the secondary, and the hybrid CR species. As the transition between the primary and the secondary CR species, it is reasonable that the hybrid species (N) have a value of $\nu_1$ between the other 2 species. Moreover, the proton have a distinct value of $\nu_1$ compared with other CR primary species. What's more interesting, $\nu_1$ of O and Si (especially O) are larger than that of others. Based on the above principle of classification, O and Si CR spectra should have the least secondary components, while all the other primary nuclei species (especially the proton) should have a considerable secondary components which could influence the spectral index obviously in this rigidity region (such as C, which has about 20 \% secondary component in its flux \cite{Genolini2018}).
Another explanation to the different $\nu_1$ values of primary CR species is the different primary source injections for them. In such case, it might straightforward ascribe the specific $\nu_1$ of proton to its charge-to-mass ratio, but it is difficult to find a universal mechanism to explain the  different $\nu_1$ values for other primary species.

In the subfigure (b) of Figure \ref{fig:spectra_params}, the uncertainties of $\nu_2$ are larger than that of $\nu_1$ because of the fewer data points with larger uncertainties in high rigidity region in the AMS-02 CR spectra. The clear 3 groups in subfigure (a) are replaced by complicated relationships. Different from the relationships of $\nu_1$ for primary species, the $\nu_2$ of Ne, Mg, and Si have even larger uncertainties and cover a large interval from that of  the primary species (proton, He, C, and O) to the secondary species (Li, Be, and B). The $\nu_2$ of N is somewhat consistent with that the primary component dominates the N spectra in high rigidity region \cite{AMS02_N}.

In the subfigure (c) of Figure \ref{fig:spectra_params}, it is shown that the values of $\Rbr$ could be divided into 2 groups: proton, He, C, Li, Be, B, and N in one group; O, Ne, Mg, and Si in another group. As the daughter species of C, N\footnote{The break here is mainly determined by its primary component \cite{AMS02_N}.}, O, Ne, Mg, and Si, the secondary species (Li, Be, and B) have similar $\Rbr$ values with C and N, but show systematically different $\Rbr$ intervals with O, Ne, Mg, and Si. For the primary CR species, it is interesting that the $\Rbr$ values of proton, He, C, and N are different from that of O, Ne, Mg, and Si.
 The above different rigidity of the breaks deny the propagation scenario which introduce an unique break in diffusion coefficient to reproduce the hardening in all these CR species.

In the subfigure (d) of Figure \ref{fig:spectra_params}, the values of $\Delta \nu$ inherit large uncertainties from $\nu_2$, especially for N, O, Ne, Mg, and Si. Same as the case of $\nu_2$, the $\Delta \nu$ values of Ne, Mg, and Si have even larger uncertainties and covers a large interval from that of the primary species (proton, He, C, and O) to the secondary species (Li, Be, and B). Additionally, the $\Delta \nu$ of Si even has a negative best-fit value.\footnote{In the Figure 5 of Ref. \cite{AMS02_Ne_Mg_Si}, AMS-02 collaboration get a positive $\Delta \nu$ for the Si spectrum, because they use a power law function with same spectral indexes and break to fit the spectra of Ne, Mg, and Si simultaneously (with different normalization factors), while we use independent power law functions to fit the spectra.}  Moreover, it shows that the $\Delta \nu$ values of some primary species whose spectra have relative smaller uncertainties (proton, He, and C) are systematically smaller than that of the secondary species (Li, Be, and B), which is the reason why AMS-02 spectra data favors a break in diffusion coefficient index rather than a break in the primary source injection (see, e.g., Refs. \cite{Genolini2017,Niu2020}). However, if we note how much flux of the parents species (mainly from C, N, and O) contribute to the daughter species (Li, Be, and B), an absolutely different conclusion can be obtained.

  The CR flux of Li has the ratio of contribution from its parents species\footnote{Of course, Ne, Mg, Si, and Fe also contribute to the flux of Li, Be, and B. Here, we just list the dominating parents species, more details can be found in Ref. \cite{Genolini2018}.} as $\mathrm{C}:\mathrm{N}:\mathrm{O} \simeq 0.931:1.203:1.672$; for Be, $\mathrm{C}:\mathrm{N}:\mathrm{O} \simeq 1.178:1.215:2.300$; for B, $\mathrm{C}:\mathrm{N}:\mathrm{O} \simeq 0.956:0.864:1.265$ (see Table IV in Ref. \cite{Genolini2018}). The contributions from N and O flux are about 2 to 3 times larger than that from C flux! At the same time, the ranges of $\Delta \nu$ of Li, Be, and B are perfectly covered by that of C, N, and O (especially N and O). All of the evidences show the parents and their daughter species have some ranges of $\Delta \nu$, and it is unnecessary to produce a ``extra hardening'' in the daughter species' spectra via introducing a break in the diffusion coefficient. As for the conclusions obtained in some previous works (such as Refs. \cite{Genolini2017,Niu2020}), it can be explained as: (i) just using the B/C ratio to check the propagation models, while the $\Delta \nu$ of C is coincidentally half of that of B; (ii) the employment of proton and helium data with small uncertainties (which also have relative smaller values of $\Delta \nu$ like that of C) dilutes the impacts of the real parents species (especially N and O).\footnote{In a recent work Ref. \cite{Yuan2020jcap}, the authors use the primary spectra of C and O to reproduce the secondary spectra of Li, Be, and B successfully without adding an extra break in the diffusion coefficient, which can be naturally explained by the above discussions.}

It is shown that not only the values of $\nu_1$, $\nu_2$ and $\Delta \nu$ are different for different primary CR species, but also the relationships of $\nu_1$ (low rigidity region) and $\nu_2$ (high rigidity region) between different primary species are obviously different (especially the CR spectra of Ne, Mg, and Si represent different properties compared with other primary species, which infers that they might come from different sources in different rigidity region).

Consequently, if the spectral hardening came from the primary source acceleration or propagation, it is necessary to introduce independent primary source injection for each of the primary CR species first. For the propagation case, independent breaks and relevant diffusion coefficient indexes are also needed to reproduce the observed spectra precisely. However, except that of the proton who has a different charge-to-mass ratio, there is no clear physical motivations that other primary CR nuclei species should have different source injections, let alone the independent breaks and indexes in the diffusion coefficients for them.

Moreover, considering the flux of B is mainly contributed by N and O rather than C (whose flux itself has about 20 \% secondary component), it is a more reasonable choice to use B/O (or Li/O, Be/O) ratio to check the propagation models.

\section{Conclusion and  Outlook}

In summary, the observed AMS-02 CR nuclei spectra show complicated relationships on the spectral indexes less and greater than the rigidity of the hardening (break) at a few hundred GV between the primary CR species, which disfavors that the spectral hardening simply comes from primary source acceleration or propagation if we adhere the principles of naturality and simplicity for our CR models.

Fortunately, the superposition of different kinds of sources could naturally reproduce all the spectral indexes ($\nu_1$ and $\nu_2$) and breaks ($\Rbr$) for different CR nuclei species with a simple and natural assumption that each kind of the sources could have an unique spectral index for all the primary source injection and have different element abundances compared with the other kind (see, e.g., Refs. \cite{Yuan2011,Yue2019,Yuan2020}). Considering the different kinds of potential CR sources (such as the different population of supernovae), the assumption represents the real astrophysical situations.
In this scenario, the values of $\nu_1$ indicate that one kind of sources dominate in this rigidity region, and the values of $\nu_2$ and $\Rbr$ are the results of superposition of the other kind of sources with different spectral indexes and element abundances which have considerable flux in this rigidity region (see, e.g., Refs. \cite{Yue2019,Yuan2020}).

Because of the small number and large uncertainties of the data points greater than the break rigidity, the fitting values of $\nu_2$ and $\Rbr$ (which is closely related to the detailed properties of the second type of sources) have large uncertainties. What's more, the systematics between different experiments (mainly from the energy calibration process) prevent precise fittings on a data collection of them covering different rigidity regions. As a result, CR nuclei spectra from one single experiment (such as DAMPE) are needed to do cross checks and reveal the properties of the different kinds of sources in future.


\section*{Acknowledgments}
 JSN would like to thank the referees’ valuable and detailed suggestions, which led to a great progress in this work.
This research was supported by the National Natural Science Foundation of China (NSFC) (No. 11947125 and No. 12005124) and the Applied Basic Research Programs of Natural Science Foundation of Shanxi Province (No. 201901D111043).


\begin{thebibliography}{53}%
\makeatletter
\providecommand \@ifxundefined [1]{%
 \@ifx{#1\undefined}
}%
\providecommand \@ifnum [1]{%
 \ifnum #1\expandafter \@firstoftwo
 \else \expandafter \@secondoftwo
 \fi
}%
\providecommand \@ifx [1]{%
 \ifx #1\expandafter \@firstoftwo
 \else \expandafter \@secondoftwo
 \fi
}%
\providecommand \natexlab [1]{#1}%
\providecommand \enquote  [1]{``#1''}%
\providecommand \bibnamefont  [1]{#1}%
\providecommand \bibfnamefont [1]{#1}%
\providecommand \citenamefont [1]{#1}%
\providecommand \href@noop [0]{\@secondoftwo}%
\providecommand \href [0]{\begingroup \@sanitize@url \@href}%
\providecommand \@href[1]{\@@startlink{#1}\@@href}%
\providecommand \@@href[1]{\endgroup#1\@@endlink}%
\providecommand \@sanitize@url [0]{\catcode `\\12\catcode `\$12\catcode
  `\&12\catcode `\#12\catcode `\^12\catcode `\_12\catcode `\%12\relax}%
\providecommand \@@startlink[1]{}%
\providecommand \@@endlink[0]{}%
\providecommand \url  [0]{\begingroup\@sanitize@url \@url }%
\providecommand \@url [1]{\endgroup\@href {#1}{\urlprefix }}%
\providecommand \urlprefix  [0]{URL }%
\providecommand \Eprint [0]{\href }%
\providecommand \doibase [0]{http://dx.doi.org/}%
\providecommand \selectlanguage [0]{\@gobble}%
\providecommand \bibinfo  [0]{\@secondoftwo}%
\providecommand \bibfield  [0]{\@secondoftwo}%
\providecommand \translation [1]{[#1]}%
\providecommand \BibitemOpen [0]{}%
\providecommand \bibitemStop [0]{}%
\providecommand \bibitemNoStop [0]{.\EOS\space}%
\providecommand \EOS [0]{\spacefactor3000\relax}%
\providecommand \BibitemShut  [1]{\csname bibitem#1\endcsname}%
\let\auto@bib@innerbib\@empty
\bibitem [{\citenamefont {{Panov}}\ \emph {et~al.}(2006)\citenamefont
  {{Panov}}, \citenamefont {{Adams}},\ and\ \citenamefont {{Ahn et
  al}}}]{ATIC2006}%
  \BibitemOpen
  \bibfield  {author} {\bibinfo {author} {\bibfnamefont {A.~D.}\ \bibnamefont
  {{Panov}}}, \bibinfo {author} {\bibfnamefont {J.~H.}\ \bibnamefont
  {{Adams}}}, \ and\ \bibinfo {author} {\bibfnamefont {H.~S.}\ \bibnamefont
  {{Ahn et al}}},\ }\href@noop {} {\bibfield  {journal} {\bibinfo  {journal}
  {ArXiv Astrophysics e-prints}\ } (\bibinfo {year} {2006})},\ \Eprint
  {http://arxiv.org/abs/astro-ph/0612377} {astro-ph/0612377} \BibitemShut
  {NoStop}%
\bibitem [{\citenamefont {{Ahn}}\ \emph {et~al.}(2010)\citenamefont {{Ahn}},
  \citenamefont {{Allison}},\ and\ \citenamefont {{Bagliesi et
  al}}}]{CREAM2010}%
  \BibitemOpen
  \bibfield  {author} {\bibinfo {author} {\bibfnamefont {H.~S.}\ \bibnamefont
  {{Ahn}}}, \bibinfo {author} {\bibfnamefont {P.}~\bibnamefont {{Allison}}}, \
  and\ \bibinfo {author} {\bibfnamefont {M.~G.}\ \bibnamefont {{Bagliesi et
  al}}},\ }\href {\doibase 10.1088/2041-8205/714/1/L89} {\bibfield  {journal}
  {\bibinfo  {journal} {\apjl}\ }\textbf {\bibinfo {volume} {714}},\ \bibinfo
  {pages} {L89} (\bibinfo {year} {2010})},\ \Eprint
  {http://arxiv.org/abs/1004.1123} {arXiv:1004.1123 [astro-ph.HE]} \BibitemShut
  {NoStop}%
\bibitem [{\citenamefont {{Adriani}}\ \emph {et~al.}(2011)\citenamefont
  {{Adriani}}, \citenamefont {{Barbarino}},\ and\ \citenamefont {{Bazilevskaya
  et al}}}]{PAMELA2011}%
  \BibitemOpen
  \bibfield  {author} {\bibinfo {author} {\bibfnamefont {O.}~\bibnamefont
  {{Adriani}}}, \bibinfo {author} {\bibfnamefont {G.~C.}\ \bibnamefont
  {{Barbarino}}}, \ and\ \bibinfo {author} {\bibfnamefont {G.~A.}\ \bibnamefont
  {{Bazilevskaya et al}}},\ }\href {\doibase 10.1126/science.1199172}
  {\bibfield  {journal} {\bibinfo  {journal} {Science}\ }\textbf {\bibinfo
  {volume} {332}},\ \bibinfo {pages} {69} (\bibinfo {year} {2011})},\ \Eprint
  {http://arxiv.org/abs/1103.4055} {arXiv:1103.4055 [astro-ph.HE]} \BibitemShut
  {NoStop}%
\bibitem [{\citenamefont {{Aguilar}}\ \emph {et~al.}(2013)\citenamefont
  {{Aguilar}}, \citenamefont {{Alberti}},\ and\ \citenamefont {{Alpat et
  al}}}]{AMS2013}%
  \BibitemOpen
  \bibfield  {author} {\bibinfo {author} {\bibfnamefont {M.}~\bibnamefont
  {{Aguilar}}}, \bibinfo {author} {\bibfnamefont {G.}~\bibnamefont
  {{Alberti}}}, \ and\ \bibinfo {author} {\bibfnamefont {B.}~\bibnamefont
  {{Alpat et al}}},\ }\href {\doibase 10.1103/PhysRevLett.110.141102}
  {\bibfield  {journal} {\bibinfo  {journal} {Phys. Rev. Lett.}\ }\textbf
  {\bibinfo {volume} {110}},\ \bibinfo {eid} {141102} (\bibinfo {year}
  {2013})}\BibitemShut {NoStop}%
\bibitem [{\citenamefont {{Aguilar}}\ \emph {et~al.}(2015)\citenamefont
  {{Aguilar}}, \citenamefont {{Aisa}},\ and\ \citenamefont {{Alpat et
  al}}}]{AMS02_proton}%
  \BibitemOpen
  \bibfield  {author} {\bibinfo {author} {\bibfnamefont {M.}~\bibnamefont
  {{Aguilar}}}, \bibinfo {author} {\bibfnamefont {D.}~\bibnamefont {{Aisa}}}, \
  and\ \bibinfo {author} {\bibfnamefont {B.}~\bibnamefont {{Alpat et al}}},\
  }\href {\doibase 10.1103/PhysRevLett.114.171103} {\bibfield  {journal}
  {\bibinfo  {journal} {Phys. Rev. Lett.}\ }\textbf {\bibinfo {volume} {114}},\
  \bibinfo {eid} {171103} (\bibinfo {year} {2015})}\BibitemShut {NoStop}%
\bibitem [{\citenamefont {Aguilar}\ \emph {et~al.}(2017)\citenamefont
  {Aguilar}, \citenamefont {Ali~Cavasonza},\ and\ \citenamefont {Alpat~et
  al}}]{AMS02_He_C_O}%
  \BibitemOpen
  \bibfield  {author} {\bibinfo {author} {\bibfnamefont {M.}~\bibnamefont
  {Aguilar}}, \bibinfo {author} {\bibfnamefont {L.}~\bibnamefont
  {Ali~Cavasonza}}, \ and\ \bibinfo {author} {\bibfnamefont {B.}~\bibnamefont
  {Alpat~et al}} (\bibinfo {collaboration} {AMS Collaboration}),\ }\href
  {\doibase 10.1103/PhysRevLett.119.251101} {\bibfield  {journal} {\bibinfo
  {journal} {Phys. Rev. Lett.}\ }\textbf {\bibinfo {volume} {119}},\ \bibinfo
  {pages} {251101} (\bibinfo {year} {2017})}\BibitemShut {NoStop}%
\bibitem [{\citenamefont {{Aguilar}}\ \emph {et~al.}(2020)\citenamefont
  {{Aguilar}}, \citenamefont {{Ali Cavasonza}},\ and\ \citenamefont {{Ambrosi
  et al}}}]{AMS02_Ne_Mg_Si}%
  \BibitemOpen
  \bibfield  {author} {\bibinfo {author} {\bibfnamefont {M.}~\bibnamefont
  {{Aguilar}}}, \bibinfo {author} {\bibfnamefont {L.}~\bibnamefont {{Ali
  Cavasonza}}}, \ and\ \bibinfo {author} {\bibfnamefont {G.}~\bibnamefont
  {{Ambrosi et al}}},\ }\href {\doibase 10.1103/PhysRevLett.124.211102}
  {\bibfield  {journal} {\bibinfo  {journal} {\prl}\ }\textbf {\bibinfo
  {volume} {124}},\ \bibinfo {eid} {211102} (\bibinfo {year}
  {2020})}\BibitemShut {NoStop}%
\bibitem [{\citenamefont {Aguilar}\ \emph
  {et~al.}(2018{\natexlab{a}})\citenamefont {Aguilar}, \citenamefont
  {Ali~Cavasonza},\ and\ \citenamefont {Ambrosi~et al}}]{AMS02_Li_Be_B}%
  \BibitemOpen
  \bibfield  {author} {\bibinfo {author} {\bibfnamefont {M.}~\bibnamefont
  {Aguilar}}, \bibinfo {author} {\bibfnamefont {L.}~\bibnamefont
  {Ali~Cavasonza}}, \ and\ \bibinfo {author} {\bibfnamefont {G.}~\bibnamefont
  {Ambrosi~et al}} (\bibinfo {collaboration} {AMS Collaboration}),\ }\href
  {\doibase 10.1103/PhysRevLett.120.021101} {\bibfield  {journal} {\bibinfo
  {journal} {Phys. Rev. Lett.}\ }\textbf {\bibinfo {volume} {120}},\ \bibinfo
  {pages} {021101} (\bibinfo {year} {2018}{\natexlab{a}})}\BibitemShut
  {NoStop}%
\bibitem [{\citenamefont {Aguilar}\ \emph
  {et~al.}(2018{\natexlab{b}})\citenamefont {Aguilar}, \citenamefont
  {Ali~Cavasonza},\ and\ \citenamefont {Alpat~et al}}]{AMS02_N}%
  \BibitemOpen
  \bibfield  {author} {\bibinfo {author} {\bibfnamefont {M.}~\bibnamefont
  {Aguilar}}, \bibinfo {author} {\bibfnamefont {L.}~\bibnamefont
  {Ali~Cavasonza}}, \ and\ \bibinfo {author} {\bibfnamefont {B.}~\bibnamefont
  {Alpat~et al}} (\bibinfo {collaboration} {AMS Collaboration}),\ }\href
  {\doibase 10.1103/PhysRevLett.121.051103} {\bibfield  {journal} {\bibinfo
  {journal} {Phys. Rev. Lett.}\ }\textbf {\bibinfo {volume} {121}},\ \bibinfo
  {pages} {051103} (\bibinfo {year} {2018}{\natexlab{b}})}\BibitemShut
  {NoStop}%
\bibitem [{\citenamefont {{Ohira}}\ and\ \citenamefont
  {{Ioka}}(2011)}]{Ohira2011}%
  \BibitemOpen
  \bibfield  {author} {\bibinfo {author} {\bibfnamefont {Y.}~\bibnamefont
  {{Ohira}}}\ and\ \bibinfo {author} {\bibfnamefont {K.}~\bibnamefont
  {{Ioka}}},\ }\href {\doibase 10.1088/2041-8205/729/1/L13} {\bibfield
  {journal} {\bibinfo  {journal} {\apjl}\ }\textbf {\bibinfo {volume} {729}},\
  \bibinfo {eid} {L13} (\bibinfo {year} {2011})}\BibitemShut {NoStop}%
\bibitem [{\citenamefont {{Ptuskin}}\ \emph {et~al.}(2013)\citenamefont
  {{Ptuskin}}, \citenamefont {{Zirakashvili}},\ and\ \citenamefont
  {{Seo}}}]{Ptuskin2013}%
  \BibitemOpen
  \bibfield  {author} {\bibinfo {author} {\bibfnamefont {V.}~\bibnamefont
  {{Ptuskin}}}, \bibinfo {author} {\bibfnamefont {V.}~\bibnamefont
  {{Zirakashvili}}}, \ and\ \bibinfo {author} {\bibfnamefont {E.-S.}\
  \bibnamefont {{Seo}}},\ }\href {\doibase 10.1088/0004-637X/763/1/47}
  {\bibfield  {journal} {\bibinfo  {journal} {\apj}\ }\textbf {\bibinfo
  {volume} {763}},\ \bibinfo {eid} {47} (\bibinfo {year} {2013})},\ \Eprint
  {http://arxiv.org/abs/1212.0381} {arXiv:1212.0381 [astro-ph.HE]} \BibitemShut
  {NoStop}%
\bibitem [{\citenamefont {{Korsmeier}}\ and\ \citenamefont
  {{Cuoco}}(2016)}]{Korsmeier2016}%
  \BibitemOpen
  \bibfield  {author} {\bibinfo {author} {\bibfnamefont {M.}~\bibnamefont
  {{Korsmeier}}}\ and\ \bibinfo {author} {\bibfnamefont {A.}~\bibnamefont
  {{Cuoco}}},\ }\href@noop {} {\bibfield  {journal} {\bibinfo  {journal} {ArXiv
  e-prints}\ } (\bibinfo {year} {2016})},\ \Eprint
  {http://arxiv.org/abs/1607.06093} {arXiv:1607.06093 [astro-ph.HE]}
  \BibitemShut {NoStop}%
\bibitem [{\citenamefont {{Ohira}}\ \emph {et~al.}(2016)\citenamefont
  {{Ohira}}, \citenamefont {{Kawanaka}},\ and\ \citenamefont
  {{Ioka}}}]{Ohira2016}%
  \BibitemOpen
  \bibfield  {author} {\bibinfo {author} {\bibfnamefont {Y.}~\bibnamefont
  {{Ohira}}}, \bibinfo {author} {\bibfnamefont {N.}~\bibnamefont {{Kawanaka}}},
  \ and\ \bibinfo {author} {\bibfnamefont {K.}~\bibnamefont {{Ioka}}},\ }\href
  {\doibase 10.1103/PhysRevD.93.083001} {\bibfield  {journal} {\bibinfo
  {journal} {\prd}\ }\textbf {\bibinfo {volume} {93}},\ \bibinfo {eid} {083001}
  (\bibinfo {year} {2016})},\ \Eprint {http://arxiv.org/abs/1506.01196}
  {arXiv:1506.01196 [astro-ph.HE]} \BibitemShut {NoStop}%
\bibitem [{\citenamefont {{Boschini}}\ \emph {et~al.}(2017)\citenamefont
  {{Boschini}}, \citenamefont {{Della Torre}},\ and\ \citenamefont {{Gervasi et
  al}}}]{Boschini2017}%
  \BibitemOpen
  \bibfield  {author} {\bibinfo {author} {\bibfnamefont {M.~J.}\ \bibnamefont
  {{Boschini}}}, \bibinfo {author} {\bibfnamefont {S.}~\bibnamefont {{Della
  Torre}}}, \ and\ \bibinfo {author} {\bibfnamefont {M.}~\bibnamefont {{Gervasi
  et al}}},\ }\href {\doibase 10.3847/1538-4357/aa6e4f} {\bibfield  {journal}
  {\bibinfo  {journal} {\apj}\ }\textbf {\bibinfo {volume} {840}},\ \bibinfo
  {eid} {115} (\bibinfo {year} {2017})},\ \Eprint
  {http://arxiv.org/abs/1704.06337} {arXiv:1704.06337 [astro-ph.HE]}
  \BibitemShut {NoStop}%
\bibitem [{\citenamefont {{Niu}}\ and\ \citenamefont {{Li}}(2018)}]{Niu201801}%
  \BibitemOpen
  \bibfield  {author} {\bibinfo {author} {\bibfnamefont {J.-S.}\ \bibnamefont
  {{Niu}}}\ and\ \bibinfo {author} {\bibfnamefont {T.}~\bibnamefont {{Li}}},\
  }\href {\doibase 10.1103/PhysRevD.97.023015} {\bibfield  {journal} {\bibinfo
  {journal} {\prd}\ }\textbf {\bibinfo {volume} {97}},\ \bibinfo {eid} {023015}
  (\bibinfo {year} {2018})},\ \Eprint {http://arxiv.org/abs/1705.11089}
  {arXiv:1705.11089 [astro-ph.HE]} \BibitemShut {NoStop}%
\bibitem [{\citenamefont {Niu}\ \emph {et~al.}(2019)\citenamefont {Niu},
  \citenamefont {Li},\ and\ \citenamefont {Xu}}]{Niu2019_dampe}%
  \BibitemOpen
  \bibfield  {author} {\bibinfo {author} {\bibfnamefont {J.-S.}\ \bibnamefont
  {Niu}}, \bibinfo {author} {\bibfnamefont {T.}~\bibnamefont {Li}}, \ and\
  \bibinfo {author} {\bibfnamefont {F.-Z.}\ \bibnamefont {Xu}},\ }\href
  {\doibase 10.1140/epjc/s10052-019-6625-7} {\bibfield  {journal} {\bibinfo
  {journal} {The European Physical Journal C}\ }\textbf {\bibinfo {volume}
  {79}},\ \bibinfo {pages} {125} (\bibinfo {year} {2019})}\BibitemShut
  {NoStop}%
\bibitem [{\citenamefont {{Zhu}}\ \emph {et~al.}(2018)\citenamefont {{Zhu}},
  \citenamefont {{Yuan}},\ and\ \citenamefont {{Wei}}}]{Zhu2018}%
  \BibitemOpen
  \bibfield  {author} {\bibinfo {author} {\bibfnamefont {C.-R.}\ \bibnamefont
  {{Zhu}}}, \bibinfo {author} {\bibfnamefont {Q.}~\bibnamefont {{Yuan}}}, \
  and\ \bibinfo {author} {\bibfnamefont {D.-M.}\ \bibnamefont {{Wei}}},\ }\href
  {\doibase 10.3847/1538-4357/aacff9} {\bibfield  {journal} {\bibinfo
  {journal} {\apj}\ }\textbf {\bibinfo {volume} {863}},\ \bibinfo {eid} {119}
  (\bibinfo {year} {2018})},\ \Eprint {http://arxiv.org/abs/1807.09470}
  {arXiv:1807.09470 [astro-ph.HE]} \BibitemShut {NoStop}%
\bibitem [{\citenamefont {{Niu}}\ \emph {et~al.}(2019)\citenamefont {{Niu}},
  \citenamefont {{Li}},\ and\ \citenamefont {{Xue}}}]{Niu2019}%
  \BibitemOpen
  \bibfield  {author} {\bibinfo {author} {\bibfnamefont {J.-S.}\ \bibnamefont
  {{Niu}}}, \bibinfo {author} {\bibfnamefont {T.}~\bibnamefont {{Li}}}, \ and\
  \bibinfo {author} {\bibfnamefont {H.-F.}\ \bibnamefont {{Xue}}},\ }\href
  {\doibase 10.3847/1538-4357/ab0420} {\bibfield  {journal} {\bibinfo
  {journal} {\apj}\ }\textbf {\bibinfo {volume} {873}},\ \bibinfo {eid} {77}
  (\bibinfo {year} {2019})},\ \Eprint {http://arxiv.org/abs/1810.09301}
  {arXiv:1810.09301 [astro-ph.HE]} \BibitemShut {NoStop}%
\bibitem [{\citenamefont {{Yuan}}(2019)}]{Yuan2019SCPMA}%
  \BibitemOpen
  \bibfield  {author} {\bibinfo {author} {\bibfnamefont {Q.}~\bibnamefont
  {{Yuan}}},\ }\href {\doibase 10.1007/s11433-018-9300-0} {\bibfield  {journal}
  {\bibinfo  {journal} {Science China Physics, Mechanics, and Astronomy}\
  }\textbf {\bibinfo {volume} {62}},\ \bibinfo {eid} {49511} (\bibinfo {year}
  {2019})},\ \Eprint {http://arxiv.org/abs/1805.10649} {arXiv:1805.10649
  [astro-ph.HE]} \BibitemShut {NoStop}%
\bibitem [{\citenamefont {{Niu}}\ and\ \citenamefont {{Xue}}(2020)}]{Niu2020}%
  \BibitemOpen
  \bibfield  {author} {\bibinfo {author} {\bibfnamefont {J.-S.}\ \bibnamefont
  {{Niu}}}\ and\ \bibinfo {author} {\bibfnamefont {H.-F.}\ \bibnamefont
  {{Xue}}},\ }\href {\doibase 10.1088/1475-7516/2020/01/036} {\bibfield
  {journal} {\bibinfo  {journal} {\jcap}\ }\textbf {\bibinfo {volume} {2020}},\
  \bibinfo {eid} {036} (\bibinfo {year} {2020})},\ \Eprint
  {http://arxiv.org/abs/1902.09343} {arXiv:1902.09343 [astro-ph.HE]}
  \BibitemShut {NoStop}%
\bibitem [{\citenamefont {{Boschini}}\ \emph
  {et~al.}(2020{\natexlab{a}})\citenamefont {{Boschini}}, \citenamefont {{Della
  Torre}}, \citenamefont {{Gervasi}}, \citenamefont {{Grandi}}, \citenamefont
  {{J{\o}hannesson}}, \citenamefont {{La Vacca}}, \citenamefont {{Masi}},
  \citenamefont {{Moskalenko}}, \citenamefont {{Pensotti}}, \citenamefont
  {{Porter}}, \citenamefont {{Quadrani}}, \citenamefont {{Rancoita}},
  \citenamefont {{Rozza}},\ and\ \citenamefont {{Tacconi}}}]{Boschini2020apj}%
  \BibitemOpen
  \bibfield  {author} {\bibinfo {author} {\bibfnamefont {M.~J.}\ \bibnamefont
  {{Boschini}}}, \bibinfo {author} {\bibfnamefont {S.}~\bibnamefont {{Della
  Torre}}}, \bibinfo {author} {\bibfnamefont {M.}~\bibnamefont {{Gervasi}}},
  \bibinfo {author} {\bibfnamefont {D.}~\bibnamefont {{Grandi}}}, \bibinfo
  {author} {\bibfnamefont {G.}~\bibnamefont {{J{\o}hannesson}}}, \bibinfo
  {author} {\bibfnamefont {G.}~\bibnamefont {{La Vacca}}}, \bibinfo {author}
  {\bibfnamefont {N.}~\bibnamefont {{Masi}}}, \bibinfo {author} {\bibfnamefont
  {I.~V.}\ \bibnamefont {{Moskalenko}}}, \bibinfo {author} {\bibfnamefont
  {S.}~\bibnamefont {{Pensotti}}}, \bibinfo {author} {\bibfnamefont {T.~A.}\
  \bibnamefont {{Porter}}}, \bibinfo {author} {\bibfnamefont {L.}~\bibnamefont
  {{Quadrani}}}, \bibinfo {author} {\bibfnamefont {P.~G.}\ \bibnamefont
  {{Rancoita}}}, \bibinfo {author} {\bibfnamefont {D.}~\bibnamefont {{Rozza}}},
  \ and\ \bibinfo {author} {\bibfnamefont {M.}~\bibnamefont {{Tacconi}}},\
  }\href {\doibase 10.3847/1538-4357/ab64f1} {\bibfield  {journal} {\bibinfo
  {journal} {\apj}\ }\textbf {\bibinfo {volume} {889}},\ \bibinfo {eid} {167}
  (\bibinfo {year} {2020}{\natexlab{a}})},\ \Eprint
  {http://arxiv.org/abs/1911.03108} {arXiv:1911.03108 [astro-ph.HE]}
  \BibitemShut {NoStop}%
\bibitem [{\citenamefont {{Boschini}}\ \emph
  {et~al.}(2020{\natexlab{b}})\citenamefont {{Boschini}}, \citenamefont {{Della
  Torre}}, \citenamefont {{Gervasi}}, \citenamefont {{Grandi}}, \citenamefont
  {{J{\'o}hannesson}}, \citenamefont {{La Vacca}}, \citenamefont {{Masi}},
  \citenamefont {{Moskalenko}}, \citenamefont {{Pensotti}}, \citenamefont
  {{Porter}}, \citenamefont {{Quadrani}}, \citenamefont {{Rancoita}},
  \citenamefont {{Rozza}},\ and\ \citenamefont {{Tacconi}}}]{Boschini2020apjs}%
  \BibitemOpen
  \bibfield  {author} {\bibinfo {author} {\bibfnamefont {M.~J.}\ \bibnamefont
  {{Boschini}}}, \bibinfo {author} {\bibfnamefont {S.}~\bibnamefont {{Della
  Torre}}}, \bibinfo {author} {\bibfnamefont {M.}~\bibnamefont {{Gervasi}}},
  \bibinfo {author} {\bibfnamefont {D.}~\bibnamefont {{Grandi}}}, \bibinfo
  {author} {\bibfnamefont {G.}~\bibnamefont {{J{\'o}hannesson}}}, \bibinfo
  {author} {\bibfnamefont {G.}~\bibnamefont {{La Vacca}}}, \bibinfo {author}
  {\bibfnamefont {N.}~\bibnamefont {{Masi}}}, \bibinfo {author} {\bibfnamefont
  {I.~V.}\ \bibnamefont {{Moskalenko}}}, \bibinfo {author} {\bibfnamefont
  {S.}~\bibnamefont {{Pensotti}}}, \bibinfo {author} {\bibfnamefont {T.~A.}\
  \bibnamefont {{Porter}}}, \bibinfo {author} {\bibfnamefont {L.}~\bibnamefont
  {{Quadrani}}}, \bibinfo {author} {\bibfnamefont {P.~G.}\ \bibnamefont
  {{Rancoita}}}, \bibinfo {author} {\bibfnamefont {D.}~\bibnamefont {{Rozza}}},
  \ and\ \bibinfo {author} {\bibfnamefont {M.}~\bibnamefont {{Tacconi}}},\
  }\href {\doibase 10.3847/1538-4365/aba901} {\bibfield  {journal} {\bibinfo
  {journal} {\apjs}\ }\textbf {\bibinfo {volume} {250}},\ \bibinfo {eid} {27}
  (\bibinfo {year} {2020}{\natexlab{b}})},\ \Eprint
  {http://arxiv.org/abs/2006.01337} {arXiv:2006.01337 [astro-ph.HE]}
  \BibitemShut {NoStop}%
\bibitem [{\citenamefont {{Yuan}}\ \emph
  {et~al.}(2020{\natexlab{a}})\citenamefont {{Yuan}}, \citenamefont {{Zhu}},
  \citenamefont {{Bi}},\ and\ \citenamefont {{Wei}}}]{Yuan2020jcap}%
  \BibitemOpen
  \bibfield  {author} {\bibinfo {author} {\bibfnamefont {Q.}~\bibnamefont
  {{Yuan}}}, \bibinfo {author} {\bibfnamefont {C.-R.}\ \bibnamefont {{Zhu}}},
  \bibinfo {author} {\bibfnamefont {X.-J.}\ \bibnamefont {{Bi}}}, \ and\
  \bibinfo {author} {\bibfnamefont {D.-M.}\ \bibnamefont {{Wei}}},\ }\href
  {\doibase 10.1088/1475-7516/2020/11/027} {\bibfield  {journal} {\bibinfo
  {journal} {\jcap}\ }\textbf {\bibinfo {volume} {2020}},\ \bibinfo {eid} {027}
  (\bibinfo {year} {2020}{\natexlab{a}})},\ \Eprint
  {http://arxiv.org/abs/1810.03141} {arXiv:1810.03141 [astro-ph.HE]}
  \BibitemShut {NoStop}%
\bibitem [{\citenamefont {{Blasi}}\ \emph {et~al.}(2012)\citenamefont
  {{Blasi}}, \citenamefont {{Amato}},\ and\ \citenamefont
  {{Serpico}}}]{Blasi2012}%
  \BibitemOpen
  \bibfield  {author} {\bibinfo {author} {\bibfnamefont {P.}~\bibnamefont
  {{Blasi}}}, \bibinfo {author} {\bibfnamefont {E.}~\bibnamefont {{Amato}}}, \
  and\ \bibinfo {author} {\bibfnamefont {P.~D.}\ \bibnamefont {{Serpico}}},\
  }\href {\doibase 10.1103/PhysRevLett.109.061101} {\bibfield  {journal}
  {\bibinfo  {journal} {Phys. Rev. Lett.}\ }\textbf {\bibinfo {volume} {109}},\
  \bibinfo {eid} {061101} (\bibinfo {year} {2012})},\ \Eprint
  {http://arxiv.org/abs/1207.3706} {arXiv:1207.3706 [astro-ph.HE]} \BibitemShut
  {NoStop}%
\bibitem [{\citenamefont {{Tomassetti}}(2012)}]{Tomassetti2012}%
  \BibitemOpen
  \bibfield  {author} {\bibinfo {author} {\bibfnamefont {N.}~\bibnamefont
  {{Tomassetti}}},\ }\href {\doibase 10.1088/2041-8205/752/1/L13} {\bibfield
  {journal} {\bibinfo  {journal} {\apjl}\ }\textbf {\bibinfo {volume} {752}},\
  \bibinfo {eid} {L13} (\bibinfo {year} {2012})},\ \Eprint
  {http://arxiv.org/abs/1204.4492} {arXiv:1204.4492 [astro-ph.HE]} \BibitemShut
  {NoStop}%
\bibitem [{\citenamefont
  {{Tomassetti}}(2015{\natexlab{a}})}]{Tomassetti2015apjl01}%
  \BibitemOpen
  \bibfield  {author} {\bibinfo {author} {\bibfnamefont {N.}~\bibnamefont
  {{Tomassetti}}},\ }\href {\doibase 10.1088/2041-8205/815/1/L1} {\bibfield
  {journal} {\bibinfo  {journal} {\apjl}\ }\textbf {\bibinfo {volume} {815}},\
  \bibinfo {eid} {L1} (\bibinfo {year} {2015}{\natexlab{a}})},\ \Eprint
  {http://arxiv.org/abs/1511.04460} {arXiv:1511.04460 [astro-ph.HE]}
  \BibitemShut {NoStop}%
\bibitem [{\citenamefont
  {{Tomassetti}}(2015{\natexlab{b}})}]{Tomassetti2015prd}%
  \BibitemOpen
  \bibfield  {author} {\bibinfo {author} {\bibfnamefont {N.}~\bibnamefont
  {{Tomassetti}}},\ }\href {\doibase 10.1103/PhysRevD.92.081301} {\bibfield
  {journal} {\bibinfo  {journal} {\prd}\ }\textbf {\bibinfo {volume} {92}},\
  \bibinfo {eid} {081301(R)} (\bibinfo {year} {2015}{\natexlab{b}})},\ \Eprint
  {http://arxiv.org/abs/1509.05775} {arXiv:1509.05775 [astro-ph.HE]}
  \BibitemShut {NoStop}%
\bibitem [{\citenamefont {{Feng}}\ \emph {et~al.}(2016)\citenamefont {{Feng}},
  \citenamefont {{Tomassetti}},\ and\ \citenamefont {{Oliva}}}]{Feng2016}%
  \BibitemOpen
  \bibfield  {author} {\bibinfo {author} {\bibfnamefont {J.}~\bibnamefont
  {{Feng}}}, \bibinfo {author} {\bibfnamefont {N.}~\bibnamefont
  {{Tomassetti}}}, \ and\ \bibinfo {author} {\bibfnamefont {A.}~\bibnamefont
  {{Oliva}}},\ }\href {\doibase 10.1103/PhysRevD.94.123007} {\bibfield
  {journal} {\bibinfo  {journal} {\prd}\ }\textbf {\bibinfo {volume} {94}},\
  \bibinfo {eid} {123007} (\bibinfo {year} {2016})},\ \Eprint
  {http://arxiv.org/abs/1610.06182} {arXiv:1610.06182 [astro-ph.HE]}
  \BibitemShut {NoStop}%
\bibitem [{\citenamefont {{G{\'e}nolini}}\ \emph {et~al.}(2017)\citenamefont
  {{G{\'e}nolini}}, \citenamefont {{Serpico}},\ and\ \citenamefont {{Boudaud et
  al}}}]{Genolini2017}%
  \BibitemOpen
  \bibfield  {author} {\bibinfo {author} {\bibfnamefont {Y.}~\bibnamefont
  {{G{\'e}nolini}}}, \bibinfo {author} {\bibfnamefont {P.~D.}\ \bibnamefont
  {{Serpico}}}, \ and\ \bibinfo {author} {\bibfnamefont {M.}~\bibnamefont
  {{Boudaud et al}}},\ }\href {\doibase 10.1103/PhysRevLett.119.241101}
  {\bibfield  {journal} {\bibinfo  {journal} {Phys. Rev. Lett.}\ }\textbf
  {\bibinfo {volume} {119}},\ \bibinfo {eid} {241101} (\bibinfo {year}
  {2017})}\BibitemShut {NoStop}%
\bibitem [{\citenamefont {{Jin}}\ \emph {et~al.}(2016)\citenamefont {{Jin}},
  \citenamefont {{Guo}},\ and\ \citenamefont {{Hu}}}]{Jin2016CPC}%
  \BibitemOpen
  \bibfield  {author} {\bibinfo {author} {\bibfnamefont {C.}~\bibnamefont
  {{Jin}}}, \bibinfo {author} {\bibfnamefont {Y.-Q.}\ \bibnamefont {{Guo}}}, \
  and\ \bibinfo {author} {\bibfnamefont {H.-B.}\ \bibnamefont {{Hu}}},\ }\href
  {\doibase 10.1088/1674-1137/40/1/015101} {\bibfield  {journal} {\bibinfo
  {journal} {Chinese Physics C}\ }\textbf {\bibinfo {volume} {40}},\ \bibinfo
  {eid} {015101} (\bibinfo {year} {2016})},\ \Eprint
  {http://arxiv.org/abs/1504.06903} {arXiv:1504.06903 [astro-ph.HE]}
  \BibitemShut {NoStop}%
\bibitem [{\citenamefont {{Guo}}\ and\ \citenamefont
  {{Yuan}}(2018{\natexlab{a}})}]{Guo2018cpc}%
  \BibitemOpen
  \bibfield  {author} {\bibinfo {author} {\bibfnamefont {Y.-Q.}\ \bibnamefont
  {{Guo}}}\ and\ \bibinfo {author} {\bibfnamefont {Q.}~\bibnamefont {{Yuan}}},\
  }\href {\doibase 10.1088/1674-1137/42/7/075103} {\bibfield  {journal}
  {\bibinfo  {journal} {Chinese Physics C}\ }\textbf {\bibinfo {volume} {42}},\
  \bibinfo {eid} {075103} (\bibinfo {year} {2018}{\natexlab{a}})},\ \Eprint
  {http://arxiv.org/abs/1701.07136} {arXiv:1701.07136 [astro-ph.HE]}
  \BibitemShut {NoStop}%
\bibitem [{\citenamefont {{Guo}}\ and\ \citenamefont
  {{Yuan}}(2018{\natexlab{b}})}]{Guo2018prd}%
  \BibitemOpen
  \bibfield  {author} {\bibinfo {author} {\bibfnamefont {Y.-Q.}\ \bibnamefont
  {{Guo}}}\ and\ \bibinfo {author} {\bibfnamefont {Q.}~\bibnamefont {{Yuan}}},\
  }\href {\doibase 10.1103/PhysRevD.97.063008} {\bibfield  {journal} {\bibinfo
  {journal} {\prd}\ }\textbf {\bibinfo {volume} {97}},\ \bibinfo {eid} {063008}
  (\bibinfo {year} {2018}{\natexlab{b}})},\ \Eprint
  {http://arxiv.org/abs/1801.05904} {arXiv:1801.05904 [astro-ph.HE]}
  \BibitemShut {NoStop}%
\bibitem [{\citenamefont {{Liu}}\ \emph {et~al.}(2018)\citenamefont {{Liu}},
  \citenamefont {{Yao}},\ and\ \citenamefont {{Guo}}}]{Liu2018}%
  \BibitemOpen
  \bibfield  {author} {\bibinfo {author} {\bibfnamefont {W.}~\bibnamefont
  {{Liu}}}, \bibinfo {author} {\bibfnamefont {Y.-h.}\ \bibnamefont {{Yao}}}, \
  and\ \bibinfo {author} {\bibfnamefont {Y.-Q.}\ \bibnamefont {{Guo}}},\ }\href
  {\doibase 10.3847/1538-4357/aaef39} {\bibfield  {journal} {\bibinfo
  {journal} {\apj}\ }\textbf {\bibinfo {volume} {869}},\ \bibinfo {eid} {176}
  (\bibinfo {year} {2018})},\ \Eprint {http://arxiv.org/abs/1802.03602}
  {arXiv:1802.03602 [astro-ph.HE]} \BibitemShut {NoStop}%
\bibitem [{\citenamefont {{Fornieri}}\ \emph {et~al.}(2020)\citenamefont
  {{Fornieri}}, \citenamefont {{Gaggero}}, \citenamefont {{Guberman}},
  \citenamefont {{Brahimi}},\ and\ \citenamefont {{Marcowith}}}]{Fornieri2020}%
  \BibitemOpen
  \bibfield  {author} {\bibinfo {author} {\bibfnamefont {O.}~\bibnamefont
  {{Fornieri}}}, \bibinfo {author} {\bibfnamefont {D.}~\bibnamefont
  {{Gaggero}}}, \bibinfo {author} {\bibfnamefont {D.}~\bibnamefont
  {{Guberman}}}, \bibinfo {author} {\bibfnamefont {L.}~\bibnamefont
  {{Brahimi}}}, \ and\ \bibinfo {author} {\bibfnamefont {A.}~\bibnamefont
  {{Marcowith}}},\ }\href@noop {} {\bibfield  {journal} {\bibinfo  {journal}
  {arXiv e-prints}\ ,\ \bibinfo {eid} {arXiv:2007.15321}} (\bibinfo {year}
  {2020})},\ \Eprint {http://arxiv.org/abs/2007.15321} {arXiv:2007.15321
  [astro-ph.HE]} \BibitemShut {NoStop}%
\bibitem [{\citenamefont {{Yuan}}\ \emph {et~al.}(2011)\citenamefont {{Yuan}},
  \citenamefont {{Zhang}},\ and\ \citenamefont {{Bi}}}]{Yuan2011}%
  \BibitemOpen
  \bibfield  {author} {\bibinfo {author} {\bibfnamefont {Q.}~\bibnamefont
  {{Yuan}}}, \bibinfo {author} {\bibfnamefont {B.}~\bibnamefont {{Zhang}}}, \
  and\ \bibinfo {author} {\bibfnamefont {X.-J.}\ \bibnamefont {{Bi}}},\ }\href
  {\doibase 10.1103/PhysRevD.84.043002} {\bibfield  {journal} {\bibinfo
  {journal} {\prd}\ }\textbf {\bibinfo {volume} {84}},\ \bibinfo {eid} {043002}
  (\bibinfo {year} {2011})},\ \Eprint {http://arxiv.org/abs/1104.3357}
  {arXiv:1104.3357 [astro-ph.HE]} \BibitemShut {NoStop}%
\bibitem [{\citenamefont {{Vladimirov}}\ \emph {et~al.}(2012)\citenamefont
  {{Vladimirov}}, \citenamefont {{J{\'o}hannesson}},\ and\ \citenamefont
  {{Moskalenko et al}}}]{Vladimirov2012}%
  \BibitemOpen
  \bibfield  {author} {\bibinfo {author} {\bibfnamefont {A.~E.}\ \bibnamefont
  {{Vladimirov}}}, \bibinfo {author} {\bibfnamefont {G.}~\bibnamefont
  {{J{\'o}hannesson}}}, \ and\ \bibinfo {author} {\bibfnamefont {I.~V.}\
  \bibnamefont {{Moskalenko et al}}},\ }\href {\doibase
  10.1088/0004-637X/752/1/68} {\bibfield  {journal} {\bibinfo  {journal}
  {\apj}\ }\textbf {\bibinfo {volume} {752}},\ \bibinfo {eid} {68} (\bibinfo
  {year} {2012})},\ \Eprint {http://arxiv.org/abs/1108.1023} {arXiv:1108.1023
  [astro-ph.HE]} \BibitemShut {NoStop}%
\bibitem [{\citenamefont {{Bernard}}\ \emph {et~al.}(2013)\citenamefont
  {{Bernard}}, \citenamefont {{Delahaye}},\ and\ \citenamefont {{Keum et
  al}}}]{Bernard2013}%
  \BibitemOpen
  \bibfield  {author} {\bibinfo {author} {\bibfnamefont {G.}~\bibnamefont
  {{Bernard}}}, \bibinfo {author} {\bibfnamefont {T.}~\bibnamefont
  {{Delahaye}}}, \ and\ \bibinfo {author} {\bibfnamefont {Y.-Y.}\ \bibnamefont
  {{Keum et al}}},\ }\href {\doibase 10.1051/0004-6361/201321202} {\bibfield
  {journal} {\bibinfo  {journal} {\aap}\ }\textbf {\bibinfo {volume} {555}},\
  \bibinfo {eid} {A48} (\bibinfo {year} {2013})},\ \Eprint
  {http://arxiv.org/abs/1207.4670} {arXiv:1207.4670 [astro-ph.HE]} \BibitemShut
  {NoStop}%
\bibitem [{\citenamefont {{Thoudam}}\ and\ \citenamefont
  {{H{\"o}randel}}(2013)}]{Thoudam2013}%
  \BibitemOpen
  \bibfield  {author} {\bibinfo {author} {\bibfnamefont {S.}~\bibnamefont
  {{Thoudam}}}\ and\ \bibinfo {author} {\bibfnamefont {J.~R.}\ \bibnamefont
  {{H{\"o}randel}}},\ }\href {\doibase 10.1093/mnras/stt1464} {\bibfield
  {journal} {\bibinfo  {journal} {\mnras}\ }\textbf {\bibinfo {volume} {435}},\
  \bibinfo {pages} {2532} (\bibinfo {year} {2013})},\ \Eprint
  {http://arxiv.org/abs/1304.1400} {arXiv:1304.1400 [astro-ph.HE]} \BibitemShut
  {NoStop}%
\bibitem [{\citenamefont {{Tomassetti}}\ and\ \citenamefont
  {{Donato}}(2015)}]{Tomassetti2015apjl02}%
  \BibitemOpen
  \bibfield  {author} {\bibinfo {author} {\bibfnamefont {N.}~\bibnamefont
  {{Tomassetti}}}\ and\ \bibinfo {author} {\bibfnamefont {F.}~\bibnamefont
  {{Donato}}},\ }\href {\doibase 10.1088/2041-8205/803/2/L15} {\bibfield
  {journal} {\bibinfo  {journal} {\apjl}\ }\textbf {\bibinfo {volume} {803}},\
  \bibinfo {eid} {L15} (\bibinfo {year} {2015})},\ \Eprint
  {http://arxiv.org/abs/1502.06150} {arXiv:1502.06150 [astro-ph.HE]}
  \BibitemShut {NoStop}%
\bibitem [{\citenamefont {{Kachelrie{\ss}}}\ \emph {et~al.}(2015)\citenamefont
  {{Kachelrie{\ss}}}, \citenamefont {{Neronov}},\ and\ \citenamefont
  {{Semikoz}}}]{Kachelriess2015}%
  \BibitemOpen
  \bibfield  {author} {\bibinfo {author} {\bibfnamefont {M.}~\bibnamefont
  {{Kachelrie{\ss}}}}, \bibinfo {author} {\bibfnamefont {A.}~\bibnamefont
  {{Neronov}}}, \ and\ \bibinfo {author} {\bibfnamefont {D.~V.}\ \bibnamefont
  {{Semikoz}}},\ }\href {\doibase 10.1103/PhysRevLett.115.181103} {\bibfield
  {journal} {\bibinfo  {journal} {Phys. Rev. Lett.}\ }\textbf {\bibinfo
  {volume} {115}},\ \bibinfo {eid} {181103} (\bibinfo {year} {2015})},\ \Eprint
  {http://arxiv.org/abs/1504.06472} {arXiv:1504.06472 [astro-ph.HE]}
  \BibitemShut {NoStop}%
\bibitem [{\citenamefont {{Guo}}\ \emph {et~al.}(2016)\citenamefont {{Guo}},
  \citenamefont {{Hu}},\ and\ \citenamefont {{Tian}}}]{Guo2016cpc}%
  \BibitemOpen
  \bibfield  {author} {\bibinfo {author} {\bibfnamefont {Y.-Q.}\ \bibnamefont
  {{Guo}}}, \bibinfo {author} {\bibfnamefont {H.-B.}\ \bibnamefont {{Hu}}}, \
  and\ \bibinfo {author} {\bibfnamefont {Z.}~\bibnamefont {{Tian}}},\ }\href
  {\doibase 10.1088/1674-1137/40/11/115001} {\bibfield  {journal} {\bibinfo
  {journal} {Chinese Physics C}\ }\textbf {\bibinfo {volume} {40}},\ \bibinfo
  {eid} {115001} (\bibinfo {year} {2016})},\ \Eprint
  {http://arxiv.org/abs/1412.8590} {arXiv:1412.8590 [astro-ph.HE]} \BibitemShut
  {NoStop}%
\bibitem [{\citenamefont {{Kawanaka}}\ and\ \citenamefont
  {{Yanagita}}(2018)}]{Kawanaka2018}%
  \BibitemOpen
  \bibfield  {author} {\bibinfo {author} {\bibfnamefont {N.}~\bibnamefont
  {{Kawanaka}}}\ and\ \bibinfo {author} {\bibfnamefont {S.}~\bibnamefont
  {{Yanagita}}},\ }\href {\doibase 10.1103/PhysRevLett.120.041103} {\bibfield
  {journal} {\bibinfo  {journal} {Phys. Rev. Lett.}\ }\textbf {\bibinfo
  {volume} {120}},\ \bibinfo {eid} {041103} (\bibinfo {year} {2018})},\ \Eprint
  {http://arxiv.org/abs/1707.00212} {arXiv:1707.00212 [astro-ph.HE]}
  \BibitemShut {NoStop}%
\bibitem [{\citenamefont {{Liu}}\ \emph {et~al.}(2019)\citenamefont {{Liu}},
  \citenamefont {{Guo}},\ and\ \citenamefont {{Yuan}}}]{Liu2019jcap}%
  \BibitemOpen
  \bibfield  {author} {\bibinfo {author} {\bibfnamefont {W.}~\bibnamefont
  {{Liu}}}, \bibinfo {author} {\bibfnamefont {Y.-Q.}\ \bibnamefont {{Guo}}}, \
  and\ \bibinfo {author} {\bibfnamefont {Q.}~\bibnamefont {{Yuan}}},\ }\href
  {\doibase 10.1088/1475-7516/2019/10/010} {\bibfield  {journal} {\bibinfo
  {journal} {\jcap}\ }\textbf {\bibinfo {volume} {2019}},\ \bibinfo {eid} {010}
  (\bibinfo {year} {2019})},\ \Eprint {http://arxiv.org/abs/1812.09673}
  {arXiv:1812.09673 [astro-ph.HE]} \BibitemShut {NoStop}%
\bibitem [{\citenamefont {{Qiao}}\ \emph {et~al.}(2019)\citenamefont {{Qiao}},
  \citenamefont {{Liu}}, \citenamefont {{Guo}},\ and\ \citenamefont
  {{Yuan}}}]{Qiao2019}%
  \BibitemOpen
  \bibfield  {author} {\bibinfo {author} {\bibfnamefont {B.-Q.}\ \bibnamefont
  {{Qiao}}}, \bibinfo {author} {\bibfnamefont {W.}~\bibnamefont {{Liu}}},
  \bibinfo {author} {\bibfnamefont {Y.-Q.}\ \bibnamefont {{Guo}}}, \ and\
  \bibinfo {author} {\bibfnamefont {Q.}~\bibnamefont {{Yuan}}},\ }\href
  {\doibase 10.1088/1475-7516/2019/12/007} {\bibfield  {journal} {\bibinfo
  {journal} {\jcap}\ }\textbf {\bibinfo {volume} {2019}},\ \bibinfo {eid} {007}
  (\bibinfo {year} {2019})},\ \Eprint {http://arxiv.org/abs/1905.12505}
  {arXiv:1905.12505 [astro-ph.HE]} \BibitemShut {NoStop}%
\bibitem [{\citenamefont {{Yang}}\ and\ \citenamefont
  {{Aharonian}}(2019)}]{Yang2019}%
  \BibitemOpen
  \bibfield  {author} {\bibinfo {author} {\bibfnamefont {R.}~\bibnamefont
  {{Yang}}}\ and\ \bibinfo {author} {\bibfnamefont {F.}~\bibnamefont
  {{Aharonian}}},\ }\href {\doibase 10.1103/PhysRevD.100.063020} {\bibfield
  {journal} {\bibinfo  {journal} {\prd}\ }\textbf {\bibinfo {volume} {100}},\
  \bibinfo {eid} {063020} (\bibinfo {year} {2019})},\ \Eprint
  {http://arxiv.org/abs/1812.04364} {arXiv:1812.04364 [astro-ph.HE]}
  \BibitemShut {NoStop}%
\bibitem [{\citenamefont {{Yue}}\ \emph {et~al.}(2019)\citenamefont {{Yue}},
  \citenamefont {{Ma}},\ and\ \citenamefont {{Yuan et al}}}]{Yue2019}%
  \BibitemOpen
  \bibfield  {author} {\bibinfo {author} {\bibfnamefont {C.}~\bibnamefont
  {{Yue}}}, \bibinfo {author} {\bibfnamefont {P.-X.}\ \bibnamefont {{Ma}}}, \
  and\ \bibinfo {author} {\bibfnamefont {Q.}~\bibnamefont {{Yuan et al}}},\
  }\href {\doibase 10.1007/s11467-019-0946-8} {\bibfield  {journal} {\bibinfo
  {journal} {Frontiers of Physics}\ }\textbf {\bibinfo {volume} {15}},\
  \bibinfo {eid} {24601} (\bibinfo {year} {2019})},\ \Eprint
  {http://arxiv.org/abs/1909.12857} {arXiv:1909.12857 [astro-ph.HE]}
  \BibitemShut {NoStop}%
\bibitem [{\citenamefont {{Yuan}}\ \emph
  {et~al.}(2020{\natexlab{b}})\citenamefont {{Yuan}}, \citenamefont {{Qiao}},
  \citenamefont {{Guo}}, \citenamefont {{Fan}},\ and\ \citenamefont
  {{Bi}}}]{Yuan2020}%
  \BibitemOpen
  \bibfield  {author} {\bibinfo {author} {\bibfnamefont {Q.}~\bibnamefont
  {{Yuan}}}, \bibinfo {author} {\bibfnamefont {B.-Q.}\ \bibnamefont {{Qiao}}},
  \bibinfo {author} {\bibfnamefont {Y.-Q.}\ \bibnamefont {{Guo}}}, \bibinfo
  {author} {\bibfnamefont {Y.-Z.}\ \bibnamefont {{Fan}}}, \ and\ \bibinfo
  {author} {\bibfnamefont {X.-J.}\ \bibnamefont {{Bi}}},\ }\href {\doibase
  10.1007/s11467-020-0990-4} {\bibfield  {journal} {\bibinfo  {journal}
  {Frontiers of Physics}\ }\textbf {\bibinfo {volume} {16}},\ \bibinfo {eid}
  {24501} (\bibinfo {year} {2020}{\natexlab{b}})},\ \Eprint
  {http://arxiv.org/abs/2007.01768} {arXiv:2007.01768 [astro-ph.HE]}
  \BibitemShut {NoStop}%
\bibitem [{\citenamefont {{Foreman-Mackey}}\ \emph {et~al.}(2013)\citenamefont
  {{Foreman-Mackey}}, \citenamefont {{Hogg}}, \citenamefont {{Lang}},\ and\
  \citenamefont {{Goodman}}}]{emcee}%
  \BibitemOpen
  \bibfield  {author} {\bibinfo {author} {\bibfnamefont {D.}~\bibnamefont
  {{Foreman-Mackey}}}, \bibinfo {author} {\bibfnamefont {D.~W.}\ \bibnamefont
  {{Hogg}}}, \bibinfo {author} {\bibfnamefont {D.}~\bibnamefont {{Lang}}}, \
  and\ \bibinfo {author} {\bibfnamefont {J.}~\bibnamefont {{Goodman}}},\ }\href
  {\doibase 10.1086/670067} {\bibfield  {journal} {\bibinfo  {journal} {\pasp}\
  }\textbf {\bibinfo {volume} {125}},\ \bibinfo {pages} {306} (\bibinfo {year}
  {2013})},\ \Eprint {http://arxiv.org/abs/1202.3665} {arXiv:1202.3665
  [astro-ph.IM]} \BibitemShut {NoStop}%
\bibitem [{\citenamefont {{Niu}}\ \emph {et~al.}(2018)\citenamefont {{Niu}},
  \citenamefont {{Li}}, \citenamefont {{Ding}}, \citenamefont {{Zhu}},
  \citenamefont {{Xue}},\ and\ \citenamefont {{Wang}}}]{Niu201802}%
  \BibitemOpen
  \bibfield  {author} {\bibinfo {author} {\bibfnamefont {J.-S.}\ \bibnamefont
  {{Niu}}}, \bibinfo {author} {\bibfnamefont {T.}~\bibnamefont {{Li}}},
  \bibinfo {author} {\bibfnamefont {R.}~\bibnamefont {{Ding}}}, \bibinfo
  {author} {\bibfnamefont {B.}~\bibnamefont {{Zhu}}}, \bibinfo {author}
  {\bibfnamefont {H.-F.}\ \bibnamefont {{Xue}}}, \ and\ \bibinfo {author}
  {\bibfnamefont {Y.}~\bibnamefont {{Wang}}},\ }\href {\doibase
  10.1103/PhysRevD.97.083012} {\bibfield  {journal} {\bibinfo  {journal}
  {\prd}\ }\textbf {\bibinfo {volume} {97}},\ \bibinfo {eid} {083012} (\bibinfo
  {year} {2018})},\ \Eprint {http://arxiv.org/abs/1712.00372} {arXiv:1712.00372
  [astro-ph.HE]} \BibitemShut {NoStop}%
\bibitem [{\citenamefont {{Derome}}\ \emph {et~al.}(2019)\citenamefont
  {{Derome}}, \citenamefont {{Maurin}}, \citenamefont {{Salati}}, \citenamefont
  {{Boudaud}}, \citenamefont {{G{\'e}nolini}},\ and\ \citenamefont
  {{Kunz{\'e}}}}]{Derome2019}%
  \BibitemOpen
  \bibfield  {author} {\bibinfo {author} {\bibfnamefont {L.}~\bibnamefont
  {{Derome}}}, \bibinfo {author} {\bibfnamefont {D.}~\bibnamefont {{Maurin}}},
  \bibinfo {author} {\bibfnamefont {P.}~\bibnamefont {{Salati}}}, \bibinfo
  {author} {\bibfnamefont {M.}~\bibnamefont {{Boudaud}}}, \bibinfo {author}
  {\bibfnamefont {Y.}~\bibnamefont {{G{\'e}nolini}}}, \ and\ \bibinfo {author}
  {\bibfnamefont {P.}~\bibnamefont {{Kunz{\'e}}}},\ }\href {\doibase
  10.1051/0004-6361/201935717} {\bibfield  {journal} {\bibinfo  {journal}
  {\aap}\ }\textbf {\bibinfo {volume} {627}},\ \bibinfo {eid} {A158} (\bibinfo
  {year} {2019})},\ \Eprint {http://arxiv.org/abs/1904.08210} {arXiv:1904.08210
  [astro-ph.HE]} \BibitemShut {NoStop}%
\bibitem [{\citenamefont {{Weinrich}}\ \emph {et~al.}(2020)\citenamefont
  {{Weinrich}}, \citenamefont {{G{\'e}nolini}}, \citenamefont {{Boudaud}},
  \citenamefont {{Derome}},\ and\ \citenamefont {{Maurin}}}]{Weinrich202001}%
  \BibitemOpen
  \bibfield  {author} {\bibinfo {author} {\bibfnamefont {N.}~\bibnamefont
  {{Weinrich}}}, \bibinfo {author} {\bibfnamefont {Y.}~\bibnamefont
  {{G{\'e}nolini}}}, \bibinfo {author} {\bibfnamefont {M.}~\bibnamefont
  {{Boudaud}}}, \bibinfo {author} {\bibfnamefont {L.}~\bibnamefont {{Derome}}},
  \ and\ \bibinfo {author} {\bibfnamefont {D.}~\bibnamefont {{Maurin}}},\
  }\href {\doibase 10.1051/0004-6361/202037875} {\bibfield  {journal} {\bibinfo
   {journal} {\aap}\ }\textbf {\bibinfo {volume} {639}},\ \bibinfo {eid} {A131}
  (\bibinfo {year} {2020})},\ \Eprint {http://arxiv.org/abs/2002.11406}
  {arXiv:2002.11406 [astro-ph.HE]} \BibitemShut {NoStop}%
\bibitem [{\citenamefont {{Heisig}}\ \emph {et~al.}(2020)\citenamefont
  {{Heisig}}, \citenamefont {{Korsmeier}},\ and\ \citenamefont
  {{Winkler}}}]{Heisig2020}%
  \BibitemOpen
  \bibfield  {author} {\bibinfo {author} {\bibfnamefont {J.}~\bibnamefont
  {{Heisig}}}, \bibinfo {author} {\bibfnamefont {M.}~\bibnamefont
  {{Korsmeier}}}, \ and\ \bibinfo {author} {\bibfnamefont {M.~W.}\ \bibnamefont
  {{Winkler}}},\ }\href {\doibase 10.1103/PhysRevResearch.2.043017} {\bibfield
  {journal} {\bibinfo  {journal} {Physical Review Research}\ }\textbf {\bibinfo
  {volume} {2}},\ \bibinfo {eid} {043017} (\bibinfo {year} {2020})},\ \Eprint
  {http://arxiv.org/abs/2005.04237} {arXiv:2005.04237 [astro-ph.HE]}
  \BibitemShut {NoStop}%
\bibitem [{\citenamefont {{G{\'e}nolini}}\ \emph {et~al.}(2018)\citenamefont
  {{G{\'e}nolini}}, \citenamefont {{Maurin}}, \citenamefont {{Moskalenko}},\
  and\ \citenamefont {{Unger}}}]{Genolini2018}%
  \BibitemOpen
  \bibfield  {author} {\bibinfo {author} {\bibfnamefont {Y.}~\bibnamefont
  {{G{\'e}nolini}}}, \bibinfo {author} {\bibfnamefont {D.}~\bibnamefont
  {{Maurin}}}, \bibinfo {author} {\bibfnamefont {I.~V.}\ \bibnamefont
  {{Moskalenko}}}, \ and\ \bibinfo {author} {\bibfnamefont {M.}~\bibnamefont
  {{Unger}}},\ }\href {\doibase 10.1103/PhysRevC.98.034611} {\bibfield
  {journal} {\bibinfo  {journal} {\prc}\ }\textbf {\bibinfo {volume} {98}},\
  \bibinfo {eid} {034611} (\bibinfo {year} {2018})},\ \Eprint
  {http://arxiv.org/abs/1803.04686} {arXiv:1803.04686 [astro-ph.HE]}
  \BibitemShut {NoStop}%
\end{thebibliography}

%

\section*{Appendix}

\begin{figure*}[htbp]
  \centering
  \includegraphics[width=0.43\textwidth]{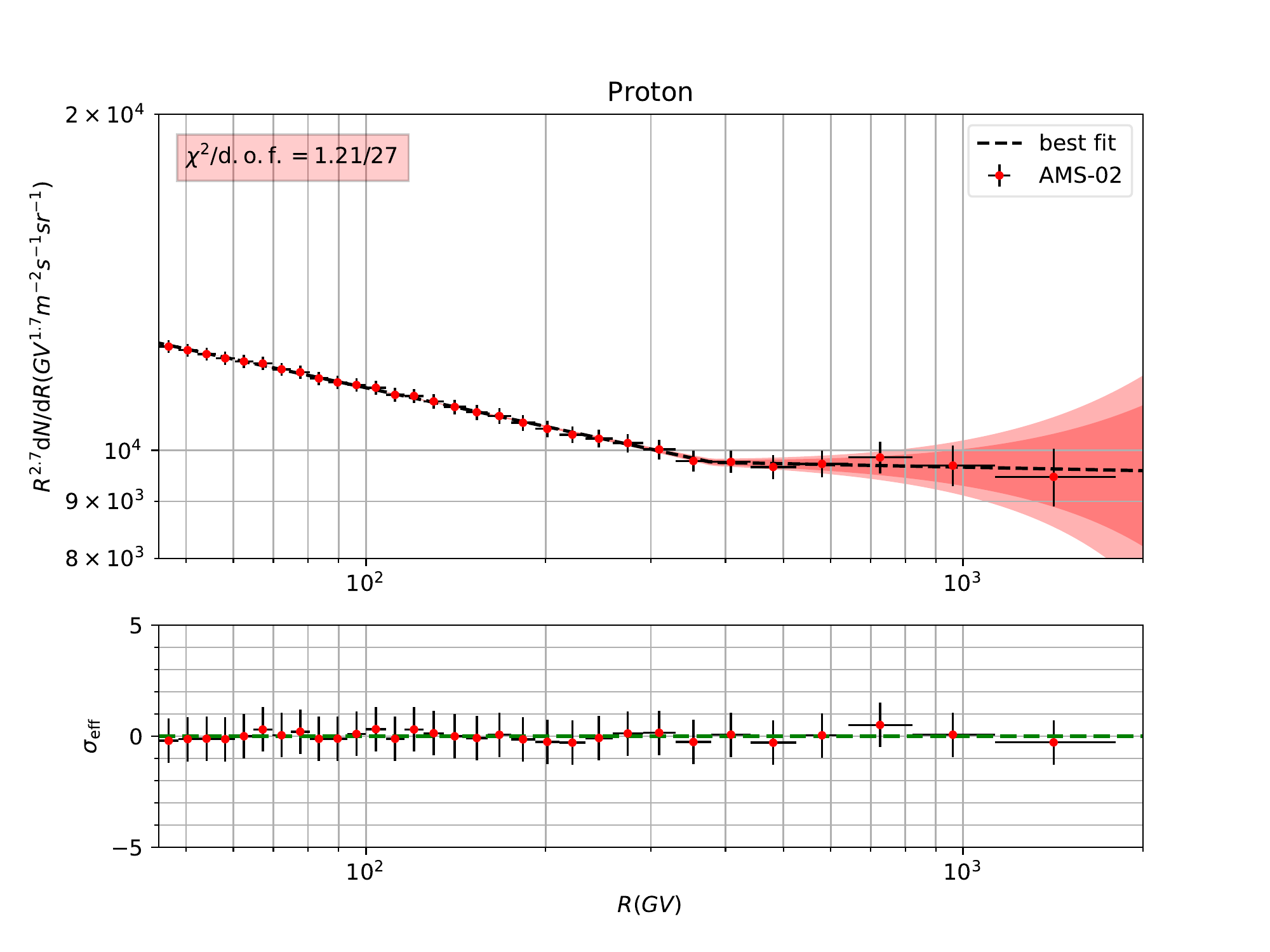}
  \includegraphics[width=0.43\textwidth]{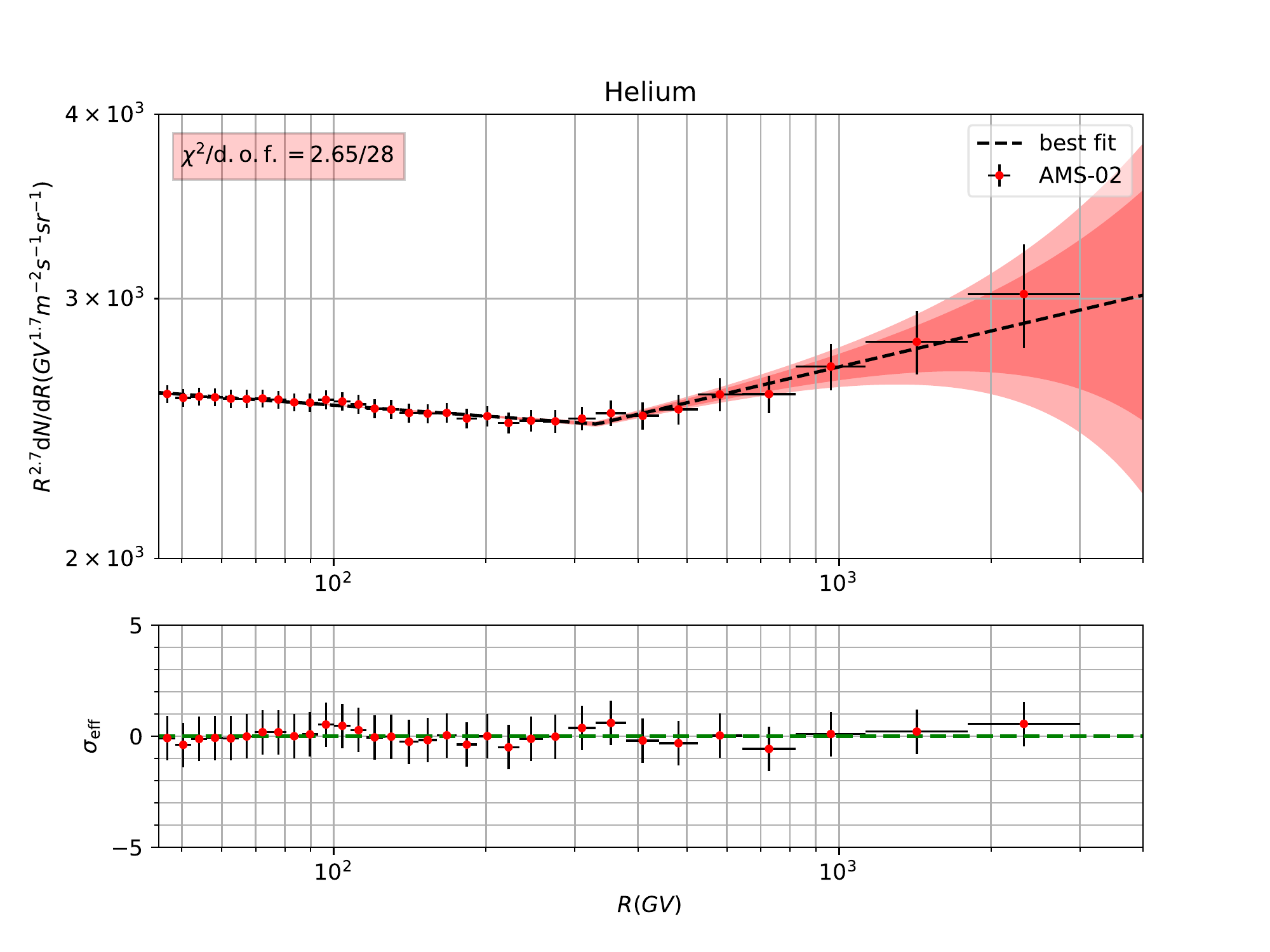}
  \includegraphics[width=0.43\textwidth]{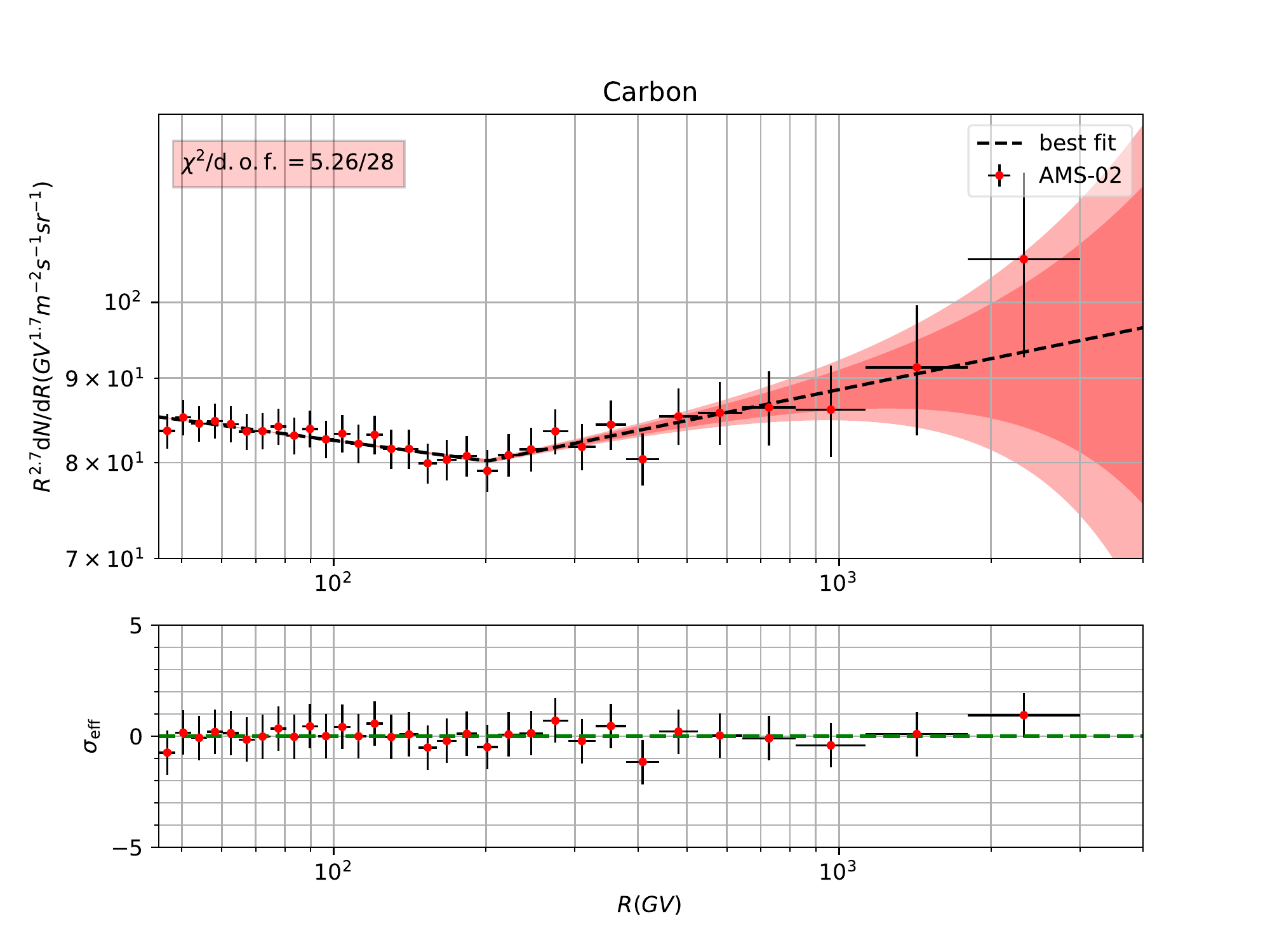}
  \includegraphics[width=0.43\textwidth]{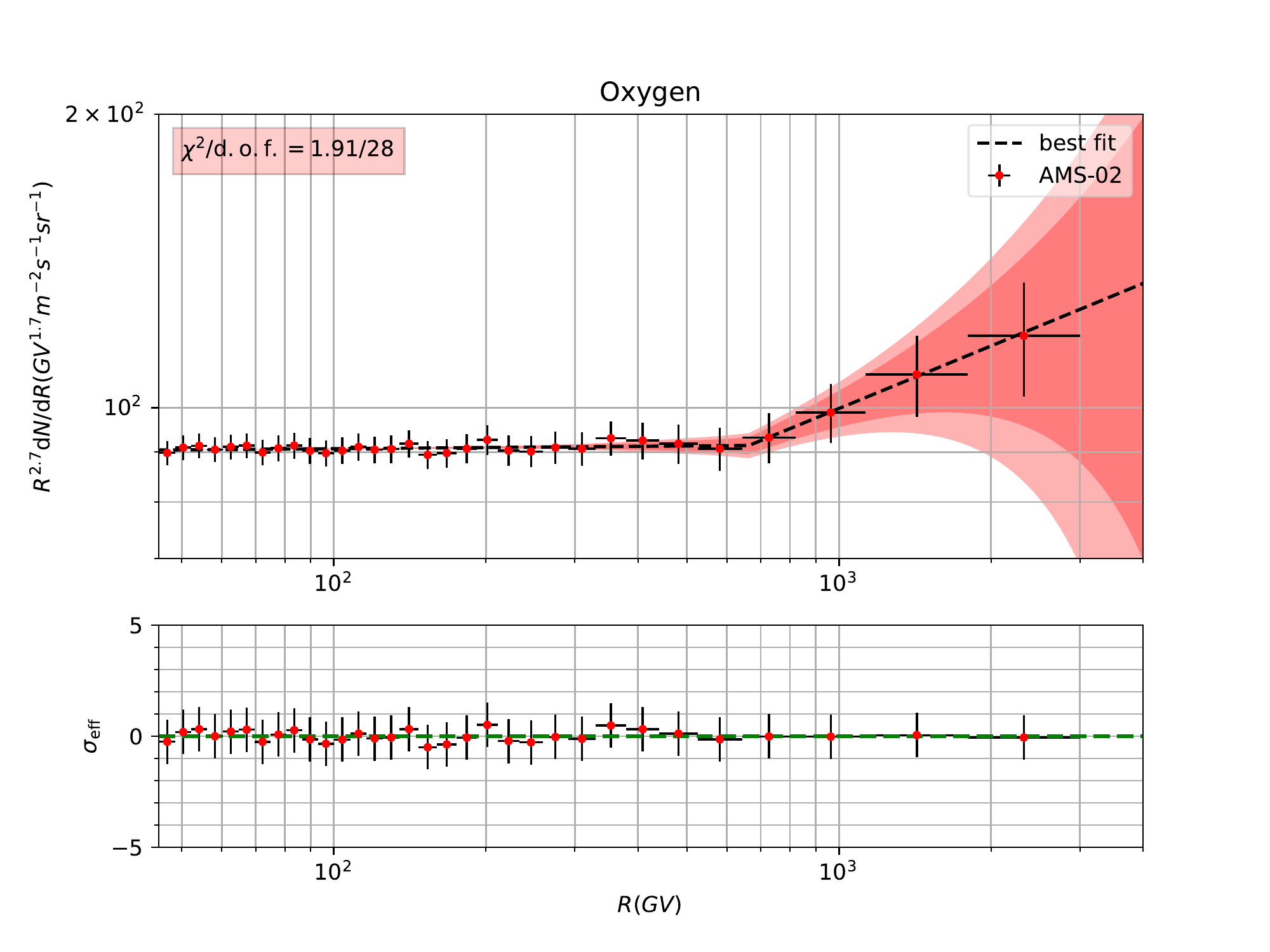}
  \includegraphics[width=0.43\textwidth]{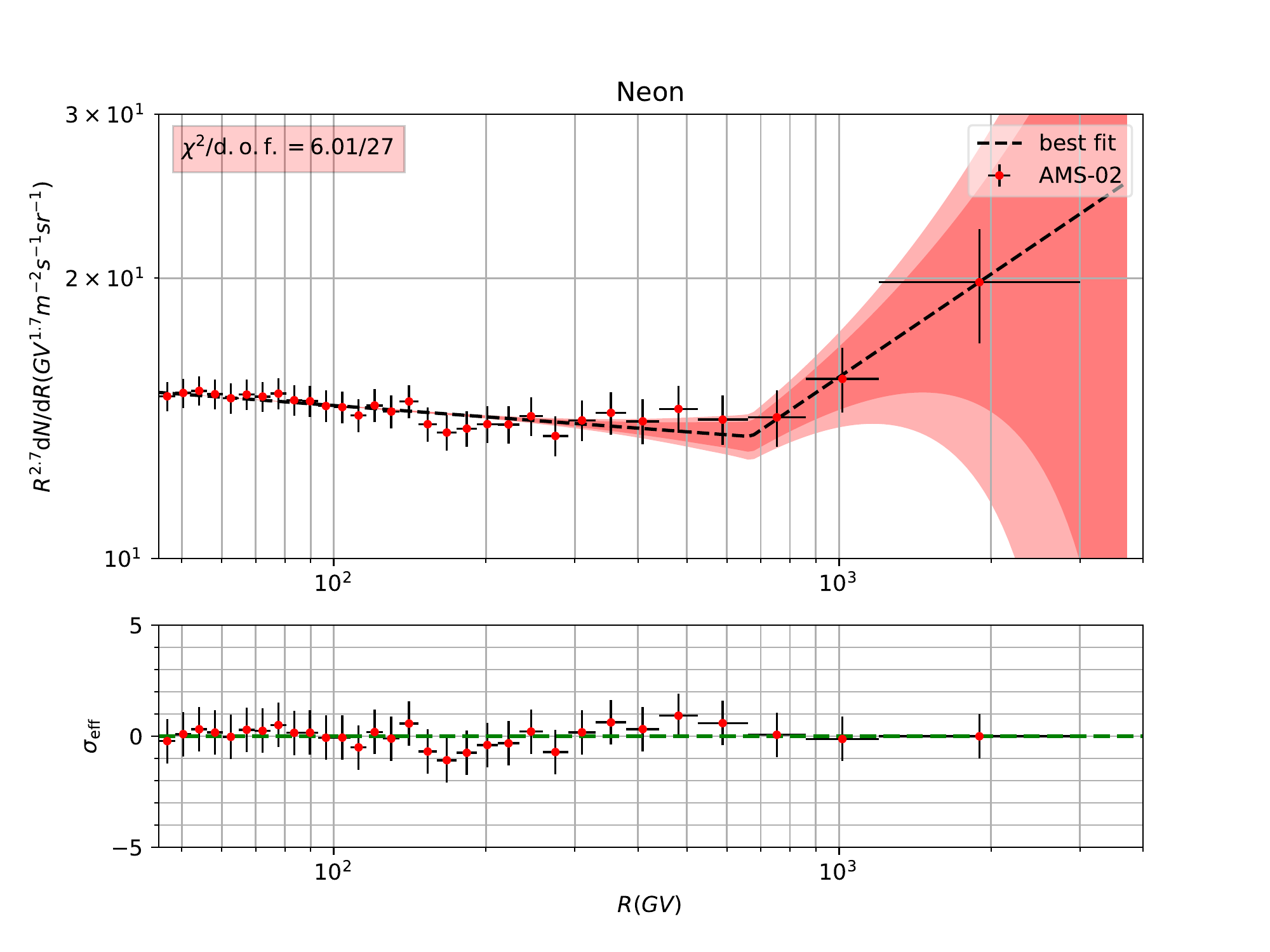}
  \includegraphics[width=0.43\textwidth]{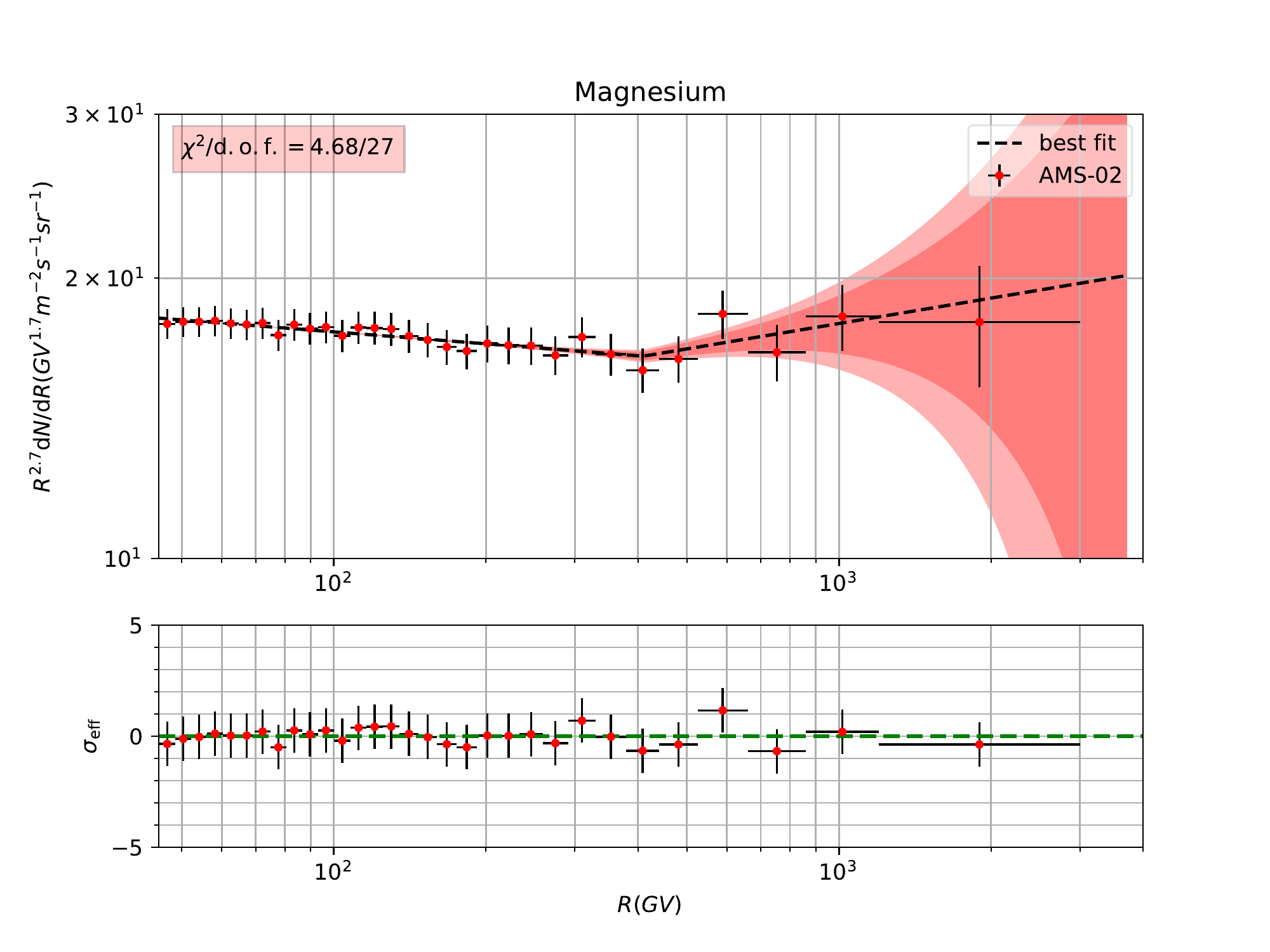}
  \includegraphics[width=0.43\textwidth]{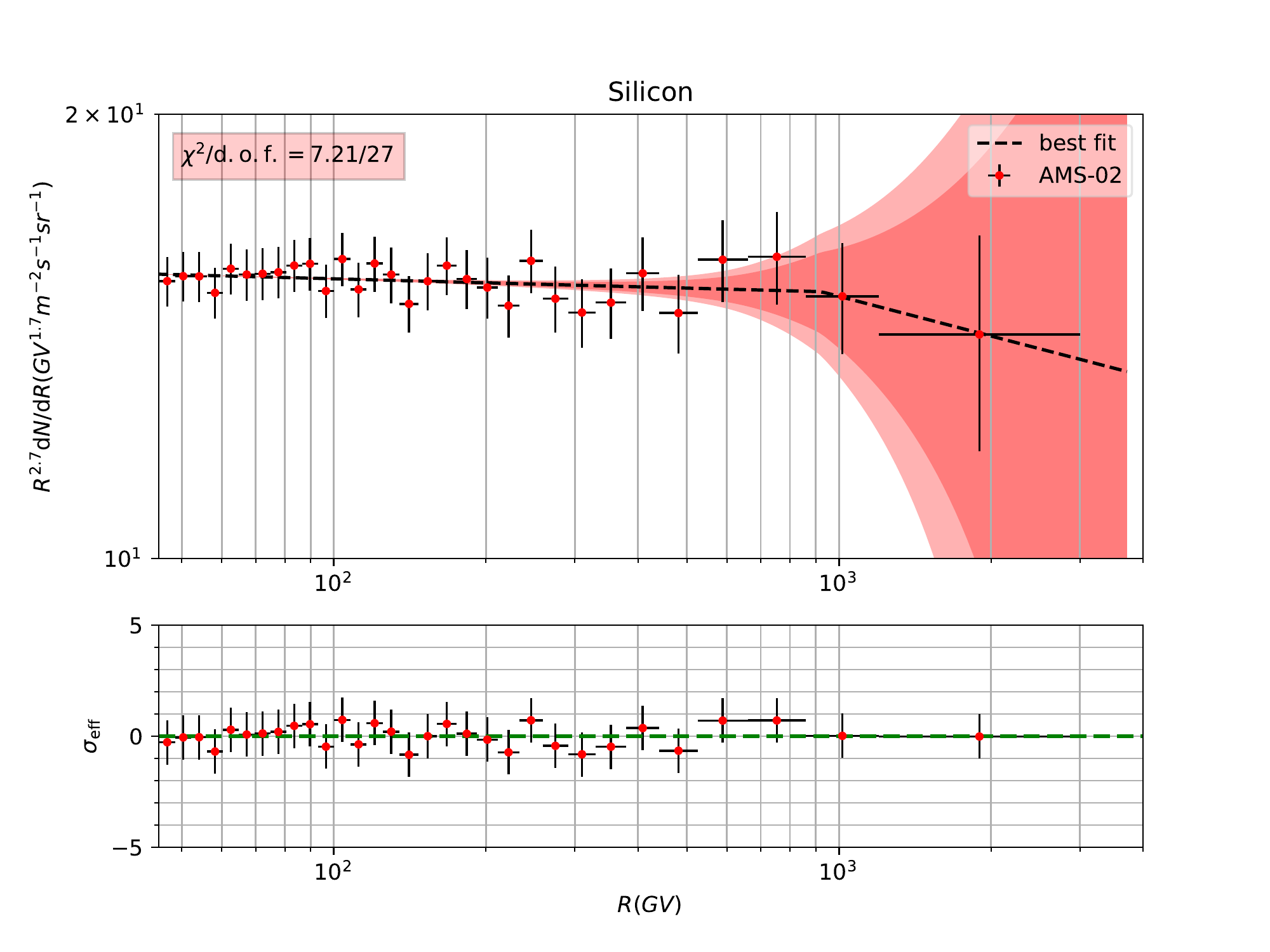}
  \caption{Fitting results and corresponding residuals to the primary CR nuclei spectra (proton, He, C, O, Ne, Mg, and Si). The 2$\sigma$ (deep red) and 3$\sigma$ (light red) bounds are also shown in the subfigures. The relevant reduced $\chi^2$ of each spectrum is given in the subfigures as well.}
  \label{fig:pri_spectra}
\end{figure*}

\begin{figure*}[htbp]
  \centering
  \includegraphics[width=0.495\textwidth]{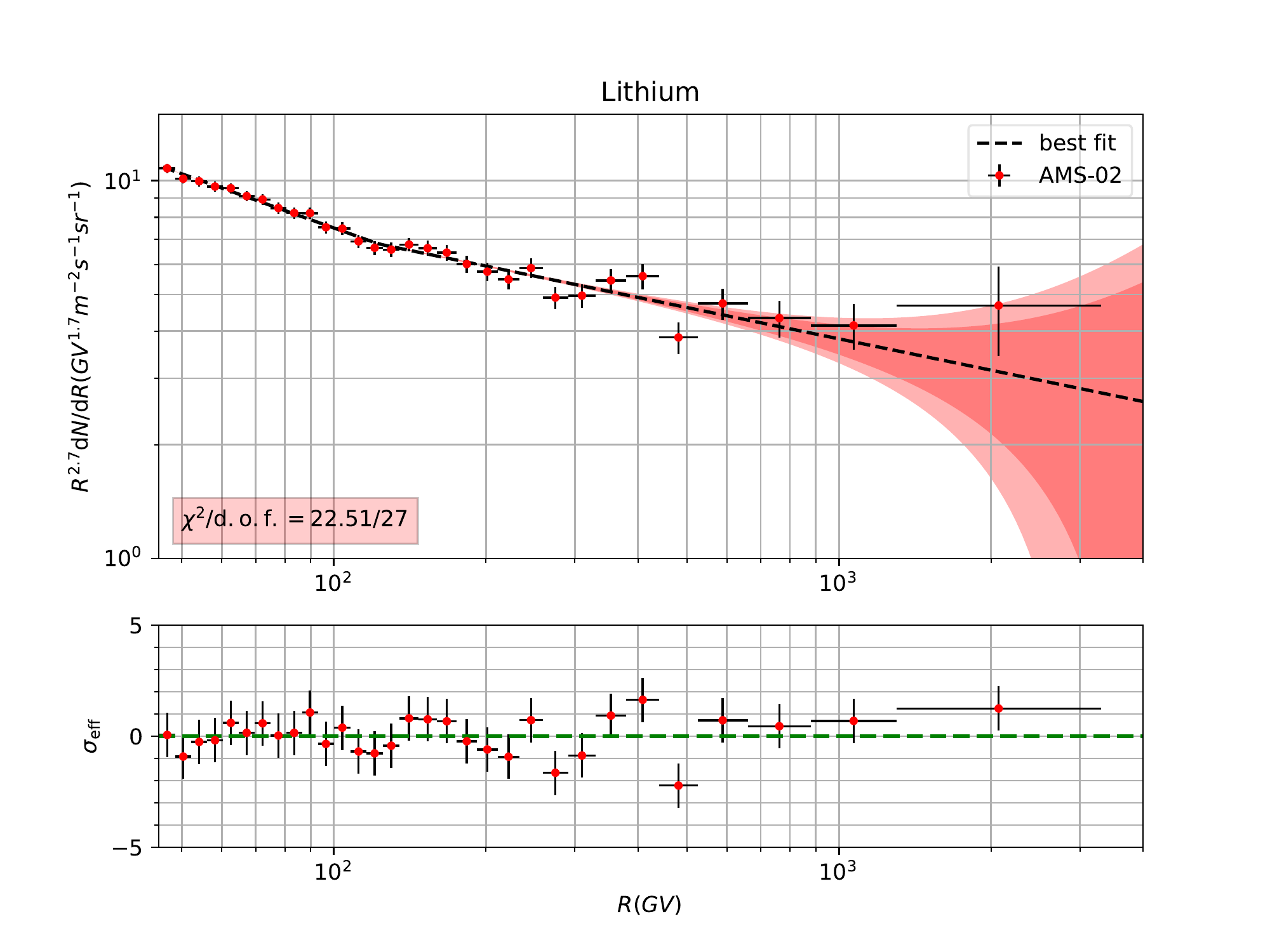}
  \includegraphics[width=0.495\textwidth]{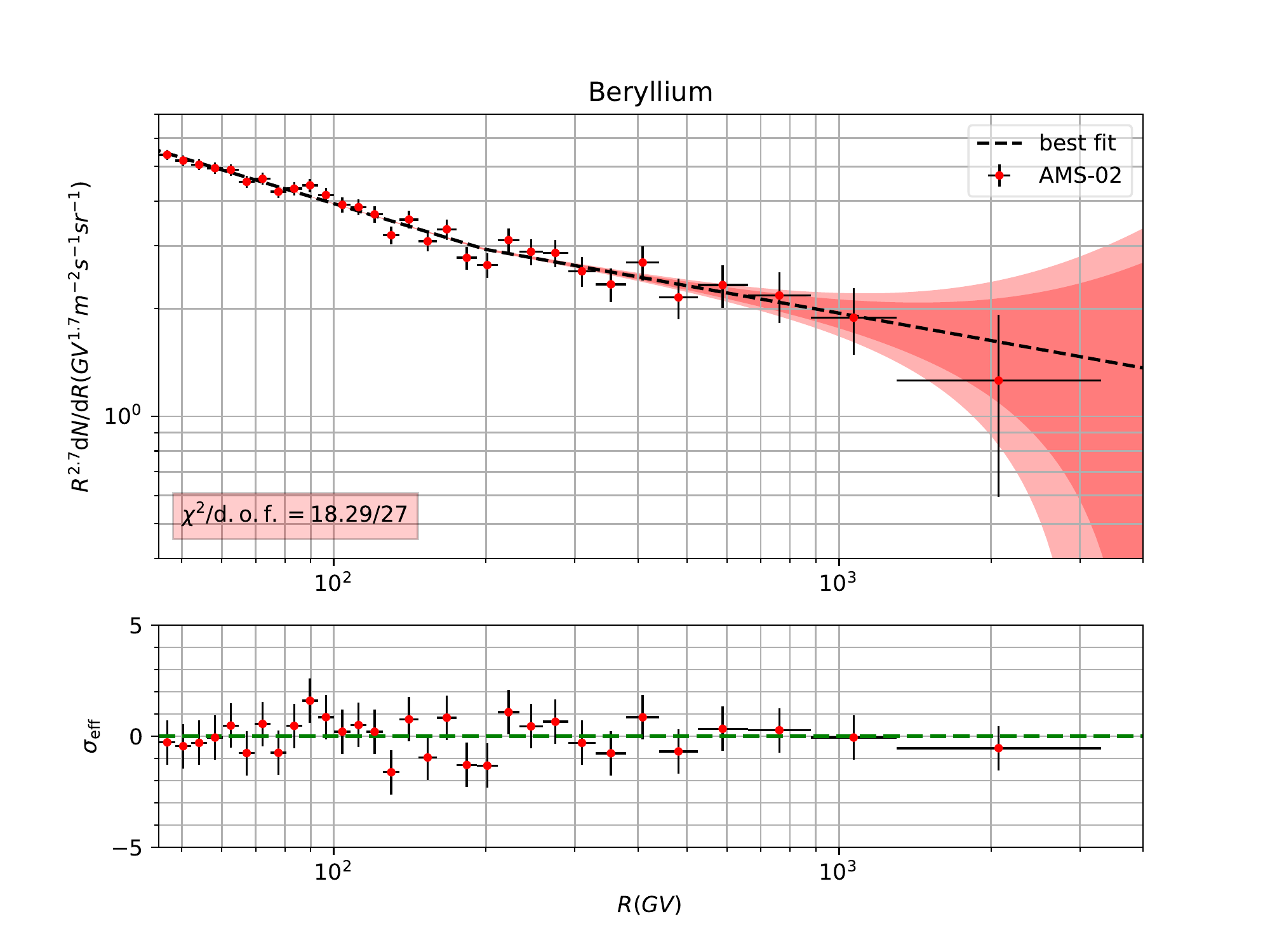}
  \includegraphics[width=0.495\textwidth]{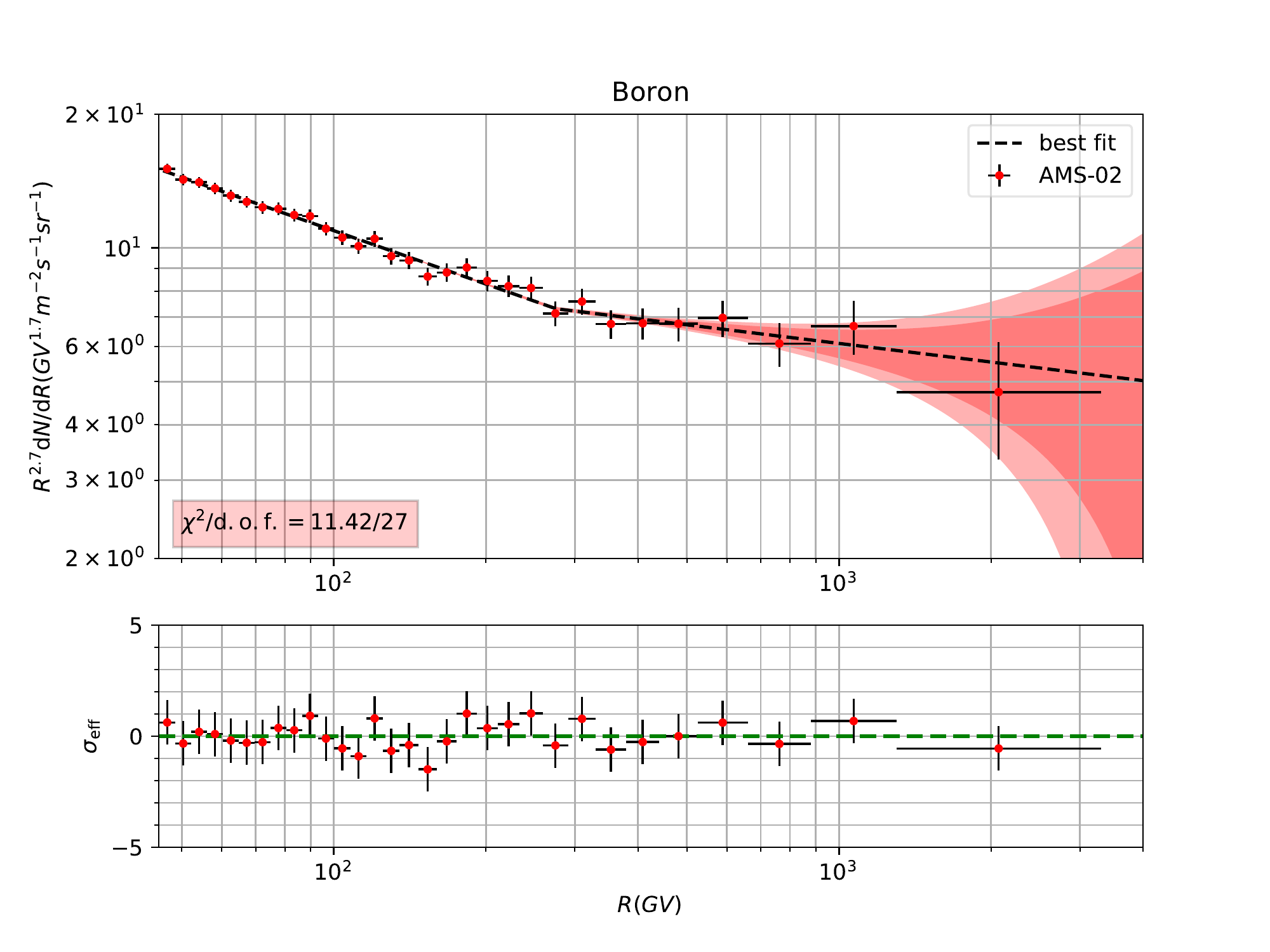}
  \caption{Fitting results and corresponding residuals to the secondary CR nuclei spectra (Li, Be, and B). The 2$\sigma$ (deep red) and 3$\sigma$ (light red) bounds are also shown in the subfigures. The relevant reduced $\chi^2$ of each spectrum is given in the subfigures as well.}
  \label{fig:sec_spectra}
\end{figure*}

\begin{figure*}[htbp]
  \centering
  \includegraphics[width=0.495\textwidth]{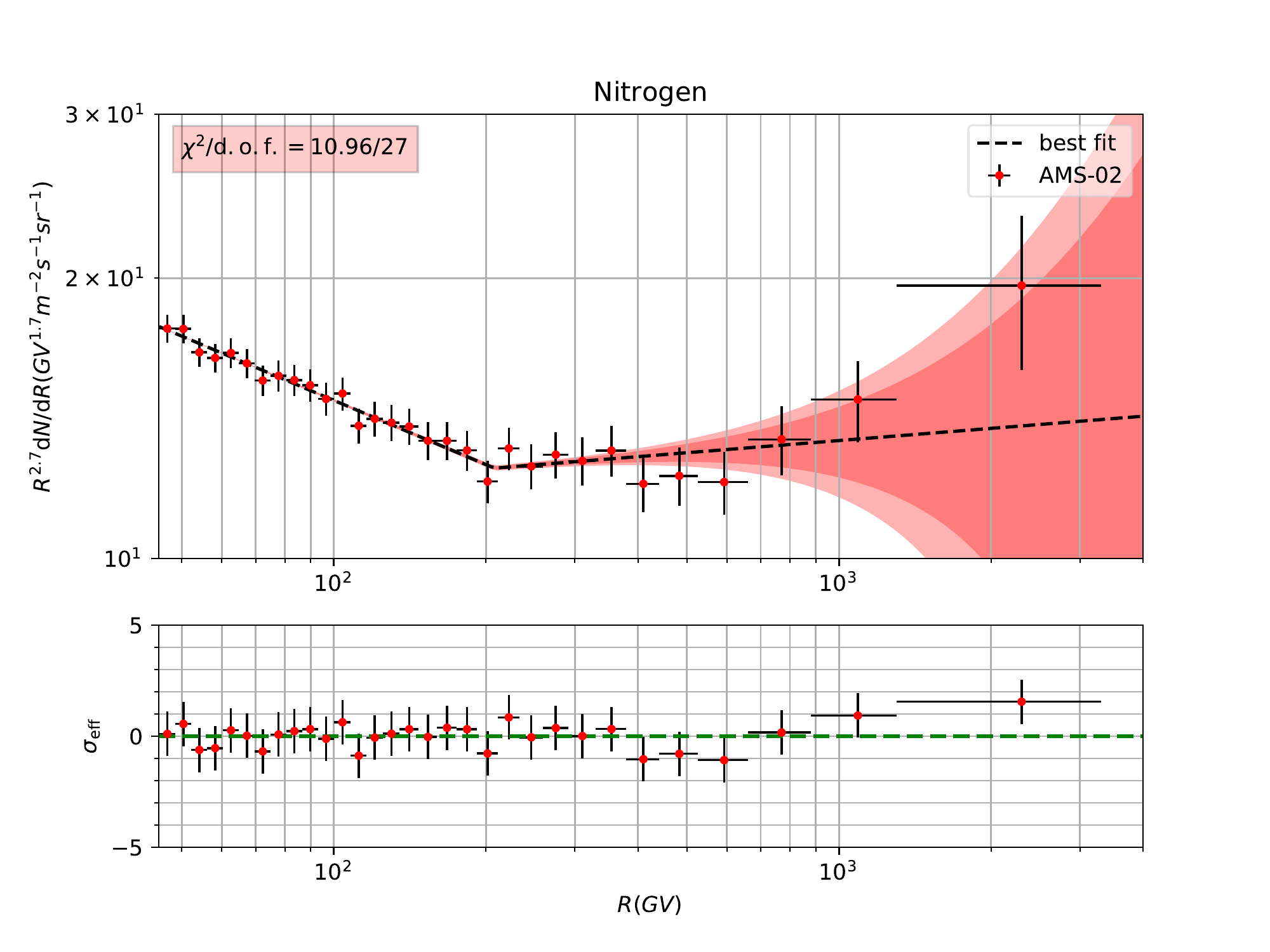}
  \caption{Fitting results and corresponding residuals to the hybrid CR nuclei spectra (N). The 2$\sigma$ (deep red) and 3$\sigma$ (light red) bounds are also shown in the subfigures. The relevant reduced $\chi^2$ of each spectrum is given in the subfigures as well.}
  \label{fig:hyb_spectra}
\end{figure*}

Note that in the lower panel of subfigures in Figs. \ref{fig:pri_spectra}, \ref{fig:sec_spectra}, and \ref{fig:hyb_spectra}, the $\sigma_{\mathrm{eff}}$ is defined as
\begin{equation}
  \sigma_{\mathrm{eff}} = \frac{f_{\mathrm{obs}} - f_\mathrm{cal}}{\sqrt{\sigma_\mathrm{stat}^{2} + \sigma_\mathrm{syst}^{2}}},
\end{equation}
where $f_\mathrm{obs}$ and $f_\mathrm{cal}$ are the points which come from the observation and model calculation; $\sigma_\mathrm{stat}$ and $\sigma_\mathrm{syst}$ are the statistical and systematic standard deviations of the observed points. This quantity could clearly show us the deviations between the best-fit result and observed values at each point based on its uncertainty.

\end{document}